\documentclass[prb,twocolumn,groupedaddress]{revtex4}
\usepackage{soul}
\usepackage{physics}

\usepackage{graphicx}

\def\presuper#1#2%
{\mathop{}%
	\mathopen{\vphantom{#2}}^{#1}%
	\kern-\scriptspace%
#2}

\usepackage{wrapfig}

\usepackage{tabularx}
\usepackage{capt-of}
\usepackage{color}
\usepackage{subfigure}
\usepackage{multirow}
\usepackage{array}
\graphicspath{{images/}{images_r2/}}
\usepackage{amsmath,amssymb,amsfonts,textcomp}
\usepackage{mathtools}

\usepackage{color}
\usepackage{subfigure}
\usepackage{multirow}
\usepackage{array}
\usepackage{outlines}    

\usepackage[normalem]{ulem}

\newcounter{ctComment}
\definecolor{dblue}{RGB}{60,105,200}
\definecolor{blue}{RGB}{99,166,159}
\definecolor{yellow}{RGB}{242,225,172}
\definecolor{orange}{RGB}{242,131,107}
\definecolor{pink}{RGB}{242,89,75}
\definecolor{red}{RGB}{205,44,36}

\begin{document}
\title{Selected Configuration Interaction in a Basis of Cluster State Tensor Products} 

\author{Vibin Abraham}
\author{Nicholas J. Mayhall}
\email{nmayhall@vt.edu}
\affiliation{Department of Chemistry, Virginia Tech,
Blacksburg, VA 24060, USA}

\begin{abstract}
	Selected configuration interaction (SCI) methods are currently enjoying a resurgence
	due to several recent developments which improve either the overall computational efficiency 
	or the compactness of the resulting SCI vector.
	These recent advances have made it possible to get full CI (FCI) quality results for much larger orbital active spaces,
	compared to conventional approaches. 
	However, due to the starting assumption that the FCI vector has 
	only a small number of significant Slater determinants,
	SCI becomes intractable for systems with strong correlation.  
	This paper introduces a method for developing SCI algorithms in a way which exploits local molecular structure to 
	significantly reduce the number of SCI variables. 
	The proposed method is defined by first grouping the orbitals into clusters over which we can define many particle cluster states.
	We then directly perform the SCI algorithm in a basis of tensor products of cluster states
	instead of Slater determinants.
	While the approach is general for arbitrarily defined cluster states, 
	we find significantly improved performance by defining cluster states through a Tucker decomposition of the 
	global (and sparse) SCI vector. 
	To demonstrate the potential of this method, called tensor product selected configuration interaction (TPSCI), 
		we present numerical results for a diverse set of examples: 1) modified Hubbard model with different inter-  and intra-cluster hopping terms,
	2) less obviously clusterable cases of bond breaking in N$_2$ and F$_2$, and 
	3) ground state energies of large planar $\pi$-conjugated systems with active spaces of up to 42 electrons in 42 orbitals.
	These numerical results show that TPSCI can be used to significantly reduce the number of SCI variables in the variational space, 
		and thus paving a path for extending these deterministic and variational SCI approaches to a wider range of physical systems.
\end{abstract}

\maketitle

\section{Introduction}

The efficient simulation of strongly correlated electrons remains a key challenge 
	toward better understanding several critical areas of chemical and molecular sciences including
	catalysis, organometallic chemistry, excited state processes, and many more. 
Although the term ``strongly correlated'' is rather ambiguously defined, 
	we will take this to mean a system which cannot be efficiently and accurately modeled using 
	perturbative or diagrammatic techniques starting from a single Slater determinant wavefunction. 
For systems dominated by one-electron interactions, Hartree-Fock (HF) represents a useful approach,
	such that the resulting single Slater determinant wavefunctions are qualitatively accurate.
Consequently, the full configuration interaction (FCI) wavefunction written as a sum of all possible determinants 
 	bears a near unit coefficient weighting the HF ground state determinant.
Methods such as perturbation theory, or coupled-cluster work extremely well in this regime.
However, as the relative strength of the two-electron component increases, 
	the HF state becomes less useful as an approximation, with many different determinants contributing significantly. 
This leads to a breakdown of most common approximations, such as perturbation theory, coupled-cluster theory, 
	density functional theory, etc.

Although any algorithm which solves an \textit{arbitrary} strongly correlated system is likely to exhibit exponential scaling,
	it is often the case that a molecule's Hamiltonian has some structure that can be exploited to make the problem easier. 
For example, if a molecule's strong correlation arises due to an orbital near-degeneracy, 
then active-space methods are effective in obtaining accurate results from a relatively simple computation. 
For systems which have a near one-dimensional structure, matrix product states provide highly efficient representations
which can be solved for using density matrix renormalization group (DMRG).\cite{White1992,Schollwock2005,Chan2011} 
Likewise, approximately two-dimensional structure can be efficiently parameterized using projected entangled pair states (PEPS),\cite{jordanClassicalSimulationInfiniteSize2008,verstraeteRenormalizationAlgorithmsQuantumMany2004}
	and recent improvements in contraction algorithms have made these algorithms more promising for molecular applications.\cite{hyattDMRGApproachOptimizing2019}
More general tensor networks have also been explored.\cite{kovyrshinSelfAdaptiveTensorNetwork2017,martiCompletegraphTensorNetwork2010,murgTreeTensorNetwork2015,szalayTensorProductMethods2015,chooFermionicNeuralnetworkStates2019}
For systems whose strong correlation occurs among a relatively small subset of Slater determinants (as opposed to a single particle subset defining an active-space)
	one might choose to perform a configuration interaction (CI) calculation using only the important Slater determinants. 
This is the physical motivation for so-called ``selected CI'' methods
	which have a long history in the field,\cite{Huron1973,Bender1969,Buenker1968} 
	but which have seen a resurgence during the past few years.\cite{Tubman2016,Schriber2016,Holmes2016,Liu2016a,Chakraborty2018,Ohtsuka2017,Coe2018,levineCASSCFExtremelyLarge2020}

Selected CI methods typically involve an iterative procedure
in which a CI calculation is performed within a small subspace of Slater determinants, 
and this subspace is iteratively refined using some search algorithm to find the most important Slater determinants. 
The basic algorithmic steps of a selected CI program involve:
\begin{enumerate}
	\item Determine an initial variational space (typically either the HF determinant or the set of single and double excitations).
	\item Find the ground state of the Hamiltonian in the current variational space.
	\item Perform some search algorithm which identifies which determinants outside of the space are most important.
		This importance criterion is usually based on perturbative or energy minimization estimates.
	\item Construct a new variational space based on the search results, and continue until the spaces stop changing.
\end{enumerate}

Although all selected CI approaches follow these general steps, they differ in various details.
As one of the first approaches of this sort, the configuration interaction
perturbatively selected iteratively (CIPSI)\cite{Huron1973} algorithm builds the CI space by adding 
determinants which have first order coefficients larger than some threshold, $\epsilon$. 
More recently, the Adaptive Sampling CI (ASCI)\cite{Tubman2016,Tubman2020,Levine2020} method accelerated the CIPSI algorithm 
	by only considering a few determinants with large coefficients in the current model space.
In that work, the authors also showed that ASCI algorithm is basically a deterministic variant of the FCI-QMC 
	proposed by Booth and coworkers.\cite{Booth2009,Cleland2010}
With the goal of achieving better accuracy guarantees, the $\Lambda$-CI method adds determinants based on a variational energy criterion 
	$\Lambda$.\cite{Evangelista2014}
Following their earlier work, Evangelista and co-workers then proposed the adaptive CI (ACI) method which produces compact wavefunctions with tunable accuracy.\cite{Schriber2016}
As a deterministic generalization of heat-bath sampling,\cite{holmesEfficientHeatBathSampling2016} 
	heat-bath CI (HCI)\cite{Holmes2016,Li2018,sharmaSemistochasticHeatBathConfiguration2017,smithCheapExactCASSCF2017} 
	adds determinants based on the magnitude of the Hamiltonian matrix element. 
This selection criteria is very cost efficient since it avoids sampling the determinants directly by using the magnitude of the integrals themselves,
	skipping the denominator computation for the selection step.
The Monte Carlo CI (MCCI) method, proposed by Greer, repeatedly adds interacting configurations randomly to the reference space and generates a variational space. \cite{Greer1995,Coe2012}

All of the SCI methods mentioned above succeed when the number of signifiant coefficients in 
	the FCI wavefunction is small, 
	and they fail when this number becomes large. 
This becomes problematic when the degree of strong correlation increases.
Luckily, the distribution of FCI wavefunction coefficients directly depends on the choice of basis. 
For example, it has been observed that rotating the one-particle basis to diagonalize an approximate one-particle density matrix (natural orbitals)
	increases the compactness of a selected CI wavefunction.\cite{Holmes2017,Tubman2016,Tubman2020}
However, single particle rotations are only the simplest type of change of basis one could consider.
This presents a natural question: \textit{Can one find a basis (not necessarily comprised of Slater determinants) 
	which yields a more compact FCI wavefunction, 
	thus decreasing the number of variational parameters in a SCI procedure? }

\begin{figure*}
	\includegraphics[width=\linewidth]{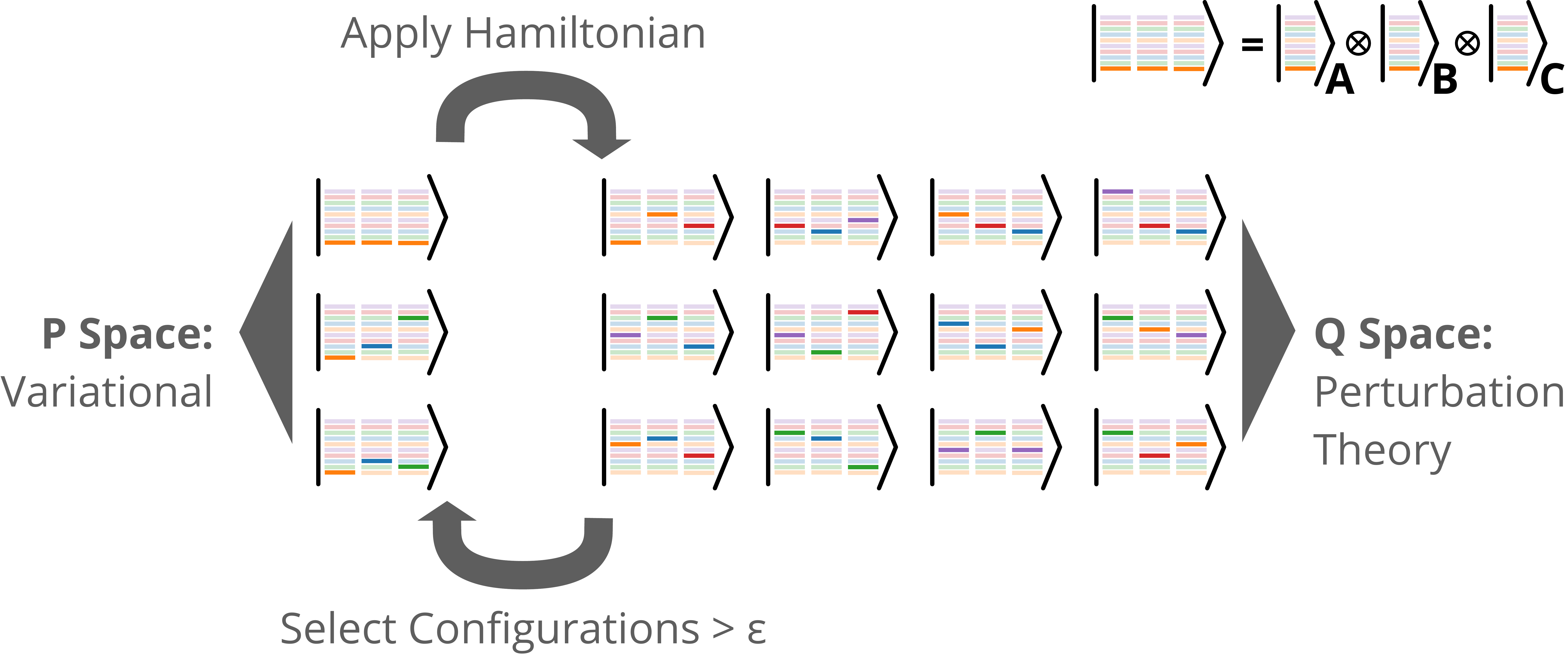}
	\caption{Schematic representation of the TPSCI algorithm for a three cluster problem.
	Each stack of lines indicates the different local states for each cluster, 
	with different colors indicating different particle number states.
	Bold colors indicate that the state is activated in that basis vector.
	The threshold, $\epsilon$, can be used to move states from $\mathcal{Q}$ to $\mathcal{P}$,
	based on the magnitude of the first order wavefunction (though other criteria could be used as well).}
	\label{fig:tpsci_algo}
\end{figure*}

In this paper, we explore a new basis designed to provide a more compact representation of the wavefunction 
	leading to larger scale SCI calculations.
This basis is defined by performing many-body rotations on disjoint sets of orbitals (or ``clusters'').
The resulting tensor product state (TPS) basis can incorporate a large amount of electron correlation into the basis vectors themselves.
As a result, the exact FCI wavefunction written in terms of tensor product states can be more compact, requiring significantly fewer 
basis vectors than in the analogous expansion in terms of Slater determinants. 
In the following sections we describe the construction of the tensor product state basis
	by way of a Tucker decomposition of a sparse global state vector,\cite{tuckerMathematicalNotesThreemode1966} 
	and a method to exploit the resulting compactness by developing a framework for 
performing CIPSI calculations directly in terms of tensor product states. 
We refer to this method as tensor product selected CI (TPSCI),
	and we investigate the numerical performance for a number of systems, 
	including the Hubbard model (Sec. \ref{sec:hubbard}), diatomic molecular dissociation (Sec. \ref{sec:diatomics}), 
	and large aromatic systems with active spaces up to 42 electrons in 42 orbitals (Sec. \ref{sec:pah}).
A cartoon schematic of the TPSCI method is shown in Fig. \ref{fig:tpsci_algo}.

%

\section{Theory}
In this section, we start by providing a description of our notation used to represent arbitrary tensor product states, 
which is similar to the work by Scuseria and co-workers in their cluster-based mean field study.\cite{Jimenez-Hoyos2015}
We then explain how the Hamiltonian matrix elements can be obtained between arbitrary tensor product states
	and give a layout of the TPSCI algorithm used in this paper.
Finally, we discuss the Tucker decomposition technique used to further compress the TPSCI wavefunction.

\subsection{Tensor Product State}

We start by partitioning the spatial orbitals into clusters.
Within each cluster, $N$, we define a set of cluster states, $\ket{\alpha}_N$, each of which is a linear combination
	of all possible Slater determinants involving a cluster's orbitals.
In order to simplify the notation, we use lower case Roman characters to enumerate orbitals ($p,q,...$), 
upper case Roman to enumerate clusters ($I,J,...$),
and Greek letters to enumerate local many-body cluster states ($\alpha, \beta, ...$).
A global tensor product state (TPS) over the full system can be represented using these cluster states as:
\begin{equation}
	\ket{\phi} = \ket{\alpha\beta\dots \gamma} = \ket{\alpha}_1 \ket{\beta}_2 ...\ket{\gamma}_N
\end{equation}
By taking all possible tensor products of local cluster states (involving all sectors of a cluster's Fock space), 
	we exactly span the original Hilbert space.
Thus, the exact full CI wavefunction can be represented in this TPS basis as
\begin{equation}
	\ket{\Psi} = \sum_{\alpha}\sum_{\beta}...\sum_{\gamma} c_{\alpha,\beta,\dots,\gamma}\ket{\alpha\beta\dots \gamma}
\end{equation}
Here $c_{\alpha,\beta,\dots,\gamma}$ is the expansion coefficient in front of the corresponding TPS configuration. 

There is freedom in how the orbitals are organized into clusters.
The orbital clustering can be chosen based on orbital locality,  symmetry of the system, or any other criterion.
While the choice of this clustering is up to the user, 
	the guiding principle is that the Hamiltonian should act more strongly within clusters and more weakly between them.
If a  reasonable clustering can be found, an accurate approximation to the FCI state can be 
	represented using few tensor product states as compared to the full space.

For the ground state of a given system, 
the tensor product state formed as the direct product of the lowest energy state of each cluster can be used as an initial approximation.
In other words, the solution for the global ground state can be approximated as
\begin{equation}
    \ket{\Psi_{0}} =  \ket{0}_1 \ket{0}_2 ... \ket{0}_n
\end{equation}
However, this approximation can be quite drastic since the cluster states are not influenced by neighboring clusters, 
	i.e., each state is the gas-phase ground state of the cluster.
Alternatively, one could choose cluster states which rigorously minimize the energy of the TPS approximation, $\ket{\Psi_{0}}$.
This approach is known as the cluster-based mean field (cMF) approximation.\cite{Jimenez-Hoyos2015}
In that work, Jimenez-Hoyos and Scuseria further minimize the energy subject to inter-cluster orbital rotations. 
The resulting cMF wavefunction, then has an optimal set of cluster states and orbitals (for representing a single TPS). 
This technique has also been recently applied to ab initio molecular systems by Hermes and Gagliardi.\cite{hermesMulticonfigurationalSelfConsistentField2019,hermesVariationalLocalizedActive2020}
While this significantly improves the reference wavefunction, only the ground state of each cluster is defined by the variational condition.
The cMF energy is invariant to rotations of the excited cluster states.
Analogous to the definition of canonical molecular orbitals in HF theory, 
the local cluster's excited state could be defined as simply the higher energy eigenvectors of the ``one-cluster reduced Hamiltonian'', 
	which is the cMF analogue of the Fock matrix.
%
Both being eigenfunctions of an effective mean-field operator, 
	the cluster state energies in cMF are analogous to the orbital energies in HF theory.
As such, one can form perturbative expansions about each mean-field operator, 
	as is done in traditional MP2 theory and in the PT2 correction developed for the cMF work.\cite{Jimenez-Hoyos2015}

\subsection{Matrix Elements}\label{sec:mat_elem}
In order to optimize the expansion coefficients of the TPS basis vectors described above, 
	one needs to evaluate the Hamiltonian matrix elements between arbitrary TPS configurations. 
Although the matrix elements in a traditional determinant-based CI code are straightforward to evaluate, 
	the TPS matrix elements are significantly more involved. 

To start, we first partition the second quantized Hamiltonian,
\begin{equation}
	\hat{H} =  \sum_{pq} h_{pq} \hat{p}^{\dagger} \hat{q} + \tfrac{1}{2}\sum_{pqrs}\mel{pq}{}{rs} 
	\hat{p}^{\dagger}\hat{q}^{\dagger}\hat{s}\hat{r}
\end{equation}
into distinct operators which are labelled by the clusters upon which they act. 
\begin{equation}\label{eq:h_terms}
\hat{H} = \sum_{I}  \hat{H}_I  + \sum_{I<J} \hat{H}_{IJ} +\sum_{I<J<K} \hat{H}_{IJK} + \sum_{I<J<K<L} \hat{H}_{IJKL} 
\end{equation}
For instance, $\hat{H}_I$ corresponds to the operators that are local to cluster, $I$.
The two-body term $\hat{H}_{IJ}$ involves all Hamiltonian operators such that the operator indices occur in  clusters $I$ and $J$.
Interactions such as 2-body charge-transfer, exchange, and dispersion fall within this set.
The details of implementing the one-body and two-body terms 
	have been worked out by Shiozaki and coworkers as part of the ASD method.\cite{parkerCommunicationActiveSpace2014,parkerCommunicationActivespaceDecomposition2013}
However, unlike in ASD which is defined for two clusters, 
	we assume an arbitrary number of clusters.
As a result, we must also handle the 3-body and 4-body Hamiltonian terms explicitly. 

Due to the antisymmetric nature of fermions, many of the above ``local'' terms 
require a non-local treatment.
For instance, to act a creation operator on cluster 3, it must first anticommute through the first two clusters.
While a general algorithm can be defined easily when using the Jordan-Wigner spin mapping
(and this was the approach we took for an initial proof of principle code), 
 this approach incurs significant overhead, and prevents efficient vectorization,
 due to the need to account for operator commutation with Pauli Z matrices. 
 To avoid this, and to speed up the matrix element construction, 
	we make the restriction that each cluster state has well-defined particle number and $\hat{S}_z$. 
 This keeps the Pauli Z strings diagonal, allowing us to simply precompute the anticommiutation sign before doing any floating point operations.  
In other words, we don't allow mixing between the local cluster states in different sectors of Fock space. 
This also has the added benefit of extra Hamiltonian sparsity and trivial quantum number preservation of the global state
(i.e., ensuring that the final state has the correct number of electrons).

To provide a concrete example, let us consider one contribution to the  two-body matrix element:
\begin{align}
	\hat{H}_{IJ} \Leftarrow& \sum_{pqr\in I}\sum_{s\in J} \mel{pq}{}{rs} 
	\hat{p}^{\dagger}\hat{q}^{\dagger}\hat{s}\hat{r} \\
	&=-\sum_{pqr\in I}\sum_{s\in J} \mel{pq}{}{rs} 
	\bigg\{\hat{p}^{\dagger}\hat{q}^{\dagger}\hat{r} \bigg\}\bigg\{\hat{s} \bigg\}
\end{align}
between two arbitrary TPS configurations: 
    \begin{align}
	    \ket{\psi} &=  \ket{\alpha}_1  ... \ket{\beta}_I ... \ket{\gamma}_J ... \ket{\delta}_N \\
	    \ket{\psi'} &=  \ket{\alpha'}_1  ... \ket{\beta'}_I ... \ket{\gamma'}_J ... \ket{\delta'}_N \\
    \end{align}


Here, we first move each group of operators to the associated cluster it acts on, keeping track of any signs.
\begin{align}
	&\sum_{pqr}^I\sum_s^J\mel{pq}{}{rs} 
	\bigg\{\hat{p}^{\dagger}\hat{q}^{\dagger}\hat{r} \bigg\}\bigg\{\hat{s} \bigg\}\ket{\alpha}_1 \dots \ket{\beta}_I \dots \ket{\gamma}_J \dots 
	\nonumber \\
	     &=(-1)^{\chi}\sum_{pqr}^{I}\sum_s^J\mel{pq}{}{rs} 
	 \ket{\alpha}_1 \dots \hat{p}^{\dagger}\hat{q}^{\dagger}\hat{r}\ket{\beta}_I \dots \hat{s} \ket{\gamma}_J\dots 
\end{align}
where $\chi = \sum_{K=I}^{J-1} N_K$, meaning we just sum the number of electrons in each state on clusters between the two active clusters.
Next applying the bra to get the matrix element, we are left with:
\begin{align}\label{eq:mel1}
	\mel{\psi'}{\hat{H}_{IJ}}{\psi} \Leftarrow& -(-1)^{\chi}\sum_{pqr\in I}\sum_{s\in J}\mel{pq}{}{rs}\Gamma_{pqr}^{\beta'\beta}\Gamma_{s}^{\gamma'\gamma} \nonumber\\
	&\times\prod_{K\neq {I,J}}\bra{\omega'}\ket{\omega}_K 
\end{align}
where $\omega$ ($\omega'$) is the local state occupied in $\ket{\psi}$ ($\ket{\psi'}$), and the operator tensor, 
\begin{align}\label{eq:gamma}
	\Gamma_{pqr}^{\beta'\beta}=\mel{\beta'}{\hat{p}^\dagger\hat{q}^\dagger\hat{r}}{\beta}_I
\end{align}
is a quantity local to cluster $I$, and 
\begin{align}\label{eq:gamma2}
	\Gamma_{s}^{\gamma'\gamma}=\mel{\gamma'}{\hat{s}}{\gamma}_J
\end{align}
is a quantity local to cluster $J$. 
In the above, all terms are zero unless the associated cluster states are from compatible cluster Hilbert spaces. 
For example, $\ket{\beta'}_I$ must have one more electron than $\ket{\beta}_I$. 
These are similar to transition density matrices, and are precomputed and accessed from memory when needed in the above expressions. 
Due to the orthonormality of the cluster states, for each cluster $K\neq \{I,J\}$, 
the overlap evaluates to a Kronecker delta indicating that the matrix element between any two TPS's is zero
	unless all of the non-active cluster states are in the same state. 
This is the TPS analogy to Slater-Condon rules, 
and significantly reduces the number of terms we must compute. 
Similar manipulations are required for all the various 1, 2, 3, and 4 body terms.

%
%

\subsection{TPSCI Algorithm}
Once the matrix elements have been implemented, then the exact state can \textit{in principle} be obtained in the TPS basis. 
However, as described in the introduction, this is intractable, and we need some method for identifying important TPS configurations.
To achieve this, we use the SCI strategies developed for Slater determinants and adapt them for generating a compact basis of TPS configurations.
While all of the different SCI strategies described above in the introduction could be leveraged in this TPS basis, 
for this initial report we simply based our work on the earlier CIPSI method, 
	often using the simplified search defined in ASCI work.\cite{Tubman2016}
A schematic overview of the TPSCI method is shown in Fig. \ref{fig:tpsci_algo}.


The algorithmic steps for TPSCI are quite similar to the Slater determinant CIPSI:
\begin{enumerate}
	\item \textbf{Precompute cluster states and operator tensors.}
		Choose a technique for defining the cluster states. 
		We find that the cMF method works well, and for many systems the orbital optimization in cMF provides significantly improved results.
		Using these defined cluster states, compute all 28 unique operator tensors, including those in Eq. \ref{eq:gamma} and \ref{eq:gamma2},
		between states in compatible sectors of the local Fock space.
		This is a very memory intensive step, limiting the approach to clusters of about 6 spatial orbitals. 
		However, if \textit{unimportant} cluster states can be identified and discarded, then 
		significant reductions in memory can be made. 
		Many approaches to this could be imagined, and we describe one technique based on an 
			approximate Schmidt decomposition in the Appendix.
		Here we also store the coefficients mapping the local Slater determinant basis to the cluster state basis.

	\item \textbf{Initialize the variational $\mathcal{P}$ space. }
		With Slater determinants, this might be done by choosing either the HF state or CISD space. 
		In this work, we initialize by deciding on an initial Fock space configuration 
		which defines how many electrons are in each cluster to begin with.
		We then choose the lowest energy TPS with that Fock space configuration. 
	 	Alternatively, one can choose multiple Fock space configurations, 
		and this is sometimes useful when describing delocalized states. 
	\item \textbf{Build the Hamiltonian matrix in the current $\mathcal{P}$ space and diagonalize. }
		As mentioned earlier, the matrix element evaluation (described in Sec. \ref{sec:mat_elem}) is more expensive than usual determinant based codes.
		Although our current code builds the full Hamiltonian matrix (limiting our current work to variational spaces containing up to around 100,000 TPS), 
		this can trivially be adapted to perform a matrix-vector product to avoid constructing the full matrix.
		This creates the current variational state, $\ket{\mathcal{P}} = \sum_i c_i \ket{\mathcal{P}_i}$,
		and variational energy, $E_0$.
	\item \textbf{Calculate the action of the Hamiltonian on each configuration in the $\mathcal{P}$ space.}
		For each TPS, $\ket{\mathcal{P}_i}$, that has a variational coefficient larger than a user-defined threshold,
		\begin{align}\label{eq:epsilon_c} 
			|c_i| > \epsilon_c,
		\end{align}
		apply each of the terms in the Hamiltonian, and collect the resulting 
		configurations that lie in the $\mathcal{Q}$ space.
		This threshold, $\epsilon_c$, is the search simplification introduced in ASCI.

		For larger systems and larger clusters, the action of the Hamiltonian on the $\mathcal{P}$ space can become too large to work with efficiently.
		Because the Hamiltonian is relatively dense in the TPS basis, 
			the action of the Hamiltonian on each configuration generates a large number of possible new configurations. 
		To increase efficiency, we introduce a screening threshold ($\epsilon_s$) to discard negligible configurations coupled by each Hamiltonian term.
		Here non-negligible is defined to be terms such that:
		\begin{align}\label{eq:epsilon_s}
			|\bra{\mathcal{Q}_j}\hat{H}_{x}\ket{\mathcal{P}_i}c_i| > \epsilon_s
		\end{align}
		where $\hat{H}_x$ is any of the terms in Eq. \ref{eq:h_terms}. 
		We then add all terms $\ket{\mathcal{Q}_j}$ where Eq. \ref{eq:epsilon_s} is {\tt true}, 
			creating a vector of coefficients in the $\mathcal{Q}$ space, $\sigma_j = \bra{\mathcal{Q}_j}\hat{H}\ket{\mathcal{P}_i}c_i$.

	\item \textbf{Compute first order coefficients in the $\mathcal{Q}$ space.}
		Using a chosen perturbative expansion (we consider either Epstein-Nesbet (EN) or the M\"oller-Plesset-like approach (MP) defined in Ref. \onlinecite{Jimenez-Hoyos2015}),
		compute the first-order coefficient for each configuration: $c^{(1)}_j = \bra{\mathcal{Q}_j}\hat{H}\ket{\mathcal{P}_i}c_i/D_j$.
		If EN partitioning is chosen,\cite{Epstein1926,Nesbet1955} 
		\begin{align}
			D_j = E_0 - \bra{\mathcal{Q}_j}\hat{H}\ket{\mathcal{Q}_j},
		\end{align}
		whereas, if MP partitioning is chosen a barycentric denominator is defined based on the effective cluster mean field operator, $\hat{F}_I$:
		\begin{align}
			D_j = \sum_I\bra{\mathcal{P}}\hat{F}_I\ket{\mathcal{P}} - \sum_I\bra{\mathcal{Q}_j}\hat{F}_I\ket{\mathcal{Q}_j},
		\end{align}
		Notice that this reduces to the MP-based approach defined for cMF if the $\mathcal{P}$ space contains only a single TPS.\cite{Jimenez-Hoyos2015}
		If cMF is used to define the cluster states, then each mean field operator, $\hat{F}_I$, is diagonal, making it computationally efficient to compute these denominators. 
		If, on the other hand, a different cluster basis is chosen (\textit{vide infra}) then we simply take the diagonal elements of $\hat{F}_I$, 
		pushing the off-diagonal elements to the perturbation.
		We have found that typically the MP partitioning is comparable to the EN partitioning, but at a much reduced cost. 

		Add any $\mathcal{Q}$ space configurations with first order coefficients larger than a threshold:
		\begin{align}
			|c^{(1)}_j|^2 > \epsilon
		\end{align}
		to the $\mathcal{P}$ space, and return to step 3, exiting if no new configurations are to be added.

\end{enumerate}

Finally, we can compute the full PT correction for the final variational ($\mathcal{P}$) space, by setting $\epsilon_c = 0$,
computing the full matrix-vector action ($\sigma_j$), and performing a dot product:
\begin{equation}
	\Delta E_{PT2}= \sum_j c_j^{(1)}\sigma_j
\end{equation}
This is usually the most expensive step in the entire calculation.

\begin{figure}
	\includegraphics[width=\linewidth]{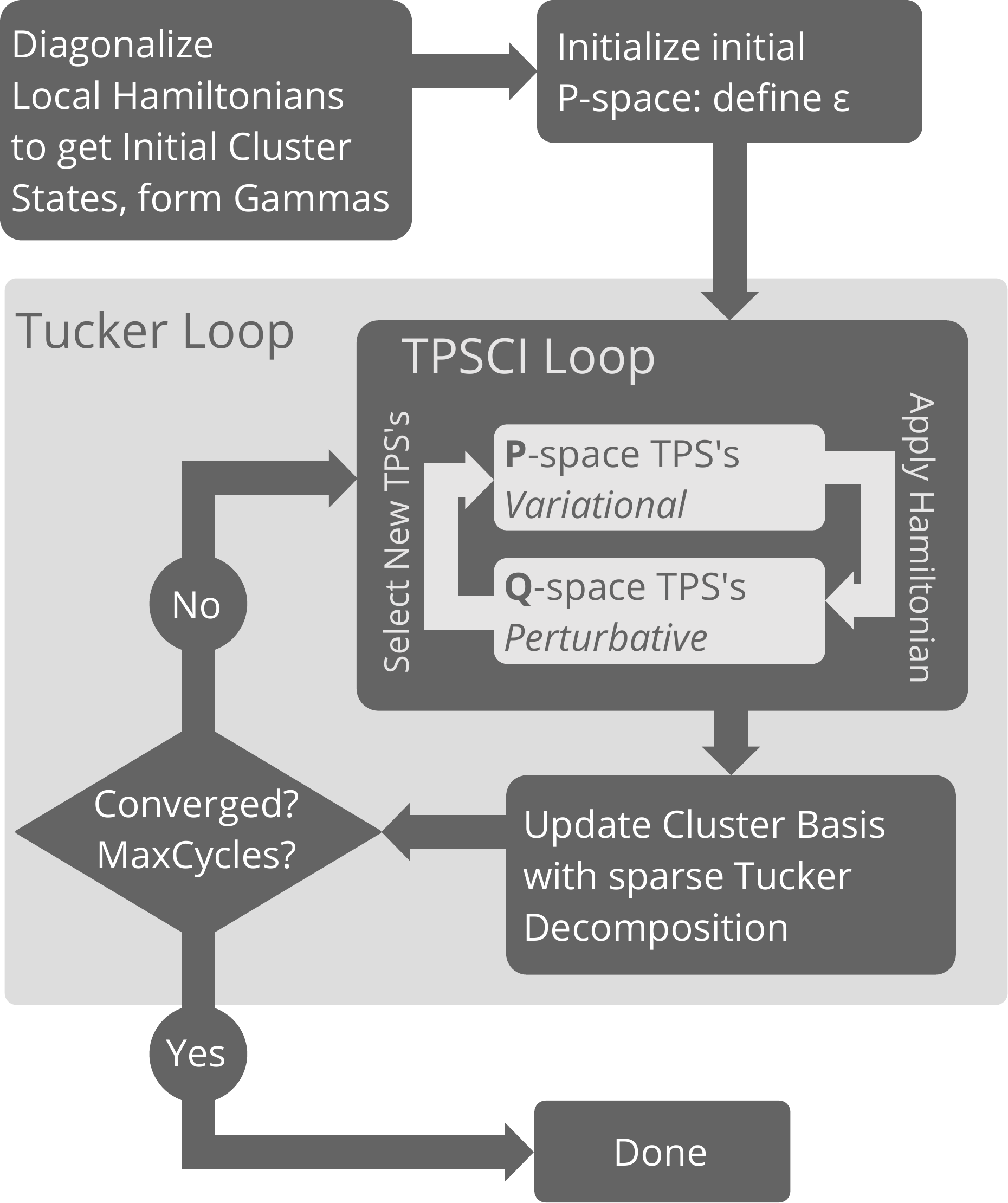}
	\caption{Self consistent Tucker decomposition loop. 
	After optimizing an approximate global state via TPSCI, 
	the sparse tensor contraction can be easily used to perform a Tucker decomposition of the state.
	This involves diagonalizing the single cluster reduced density matrices. 
	To retain local quantum numbers, we only block diagonalize the RDMs within a given Fock space. 
	Because the Tucker decomposition often significantly decreases the number of 
	variational parameters, one can optionally start with a loose threshold $\epsilon_0$ to get a better set of cluster states, 
	then tighten the threshold until it reaches the desired value, $\epsilon$, a procedure we refer to as ``bootstrapping'' explained in the  Supplementary Information.}
	\label{fig:tucker_algo}
\end{figure}

\subsection{Tucker Decomposition}
Up to this point, we have mainly considered the use of cMF for generating the cluster state basis.
This is a natural choice, as it variationally minimizes the reference TPS. 
However, this basis often does not lead to sufficiently compact representations after introducing entanglement. 
If only two clusters are present, then an SVD of the global state could be performed which would provide a 
maximally compact representation. 
When considering systems with multiple clusters, no such unique and optimal decomposition exists.
However, we can still use the same physical intuition and define the cluster state basis 
	to be the eigenvectors of the cluster's reduced density matrix. 
For two clusters, this is of course equivalent to an SVD of the global state, or a MPS. 
When the number of clusters is greater than two, this generalizes to a higher-order SVD, 
	or Tucker decomposition.\cite{tuckerMathematicalNotesThreemode1966, Mayhall2017} 
\begin{align} \label{eq:tucker_decomposition1}
	\mathcal{T}_{i,j,\dots,d} = \mathcal{C}_{\alpha,\beta,\dots,\gamma} U_{i,\alpha} U_{j,\beta} \cdots U_{d,\gamma}
\end{align}
In the absense of trunction, Eq. \ref{eq:tucker_decomposition1} is essentially a change of basis, from $i$ to $\alpha$, from $j$ to $\beta$, etc.
Each matrix $U$ can be obtained by unfolding the tensor along the associated axis and performing an SVD, e.g.,
\begin{align} 
	\mathcal{T}_{i,j\dots d} = U_{i,\alpha}\Sigma_{\alpha}V_{\alpha,j\dots d}
\end{align}
Zero values in $\Sigma$ can be dropped without approximation, revealing a subspace for the first index.
After one has formed the $U$ matrix for each index, ``core tensor'', $\mathcal{C}_{\alpha,\beta,\dots,\gamma}$, 
can be formed by a simple change of basis. 
Equivalently, the $U$ matrices can be considered as eigenvectors of the clusters reduced density matrix
evaluated via, 
\begin{align} \label{eq:tucker_decomposition}
	\rho_{i,i'} = \mathcal{T}_{i,j,\dots,d} \mathcal{T}_{i',j,\dots,d} 
\end{align}

In order to compute the Tucker decomposition of a TPSCI wavefunction, (here represented as a sparse tensor of TPS expansion coefficients) 
 we also first construct the reduced density matrix, $\rho_{\alpha'\alpha}$, sequentially for each cluster. 
Computing the reduced density matrix for cluster, $I$, involves a contraction over all clusters $J\neq I$ needs to be performed. 
\begin{align}
	\rho_{\alpha'\alpha} = c(\alpha', \beta, \dots, \gamma)c(\alpha, \beta, \dots, \gamma)
\end{align}
Fortunately, only a few of these coefficients are non-zero due to the sparsity of the TPSCI approach, 
	making the computation efficient.
Further, because we impose local quantum numbers $\hat{N}$ and  $\hat{S}_z$, 
	only the symmetry sublocks of the reduced density matrices are actually needed (with the rest being zero),
\begin{align}
	\rho_{\alpha'\alpha}^{N_\uparrow^1,N_\downarrow^1} = \tt c(\alpha'^{N_\uparrow^1,N_\downarrow^1}, \beta^{N_\uparrow^2,N_\downarrow^2}, \dots)c(\alpha^{N_\uparrow^1,N_\downarrow^1}, \beta^{N_\uparrow^2,N_\downarrow^2}, \dots)
\end{align}
such that $\alpha$ and $\alpha'$ have the same number of up and down electrons.
Once the cluster reduced density matrices are diagonalized producing new cluster states, we repeat the TPSCI calculation to get a new global vector. 
This process can be iterated until the density matrix stops changing, establishing a self-consistency condition. 
Although each iteration increases the computational cost, we find that the first iteration usually provides the most significant compression,
with subsequent iterations only making smaller changes. 
Because of this, we have found it effective to perform just a single Tucker iteration, or to use a Tucker basis from a loose TPSCI calculation (i.e., large $\epsilon$)
as the basis for tighter TPSCI calculations in a ``bootstrapping'' fashion (this is explained further
in the Supplementary Information). 
We have found that this works quite well for obtaining a compact basis at reasonable cost.
This procedure is illustrated in Fig. \ref{fig:tucker_algo}.

If the full self-consistent Tucker solution is desired, we find that the convergence is often quite quick such that the energy stops changing significantly after only a few iterations.
However, it is conceivable that this highly non-linear optimization could become a problem.
In our previous work,\cite{Mayhall2017}
	we found that a DIIS accelerated procedure which simultaneously extrapolates each cluster RDM greatly improved convergence 
	in challenging cases. 
The same strategy could be applied here if needed.

\subsection{Implementation Details}
Our current code is written in Python, using NumPy for the computation of the matrix elements.
Our code uses PySCF\cite{Sun2018} for performing RHF calculations and computing the one- and two-electron integrals.
As a pilot implementation, the performance of our code is far from optimal, 
	with many opportunities existing for optimization.
However, despite this, we are still able to perform rather non-trivial calculations on large active spaces,
	sometimes with higher accuracy than we were able to achieve with HCI or ASCI. 
We expect that an implementation in a lower-level language like {\tt C++} will increase the method's performance considerably.

One aspect of this work which is likely to impact performance is the technique chosen for storing the wavefunction. 
Since our states are sparse without any predictable structure in the indices, 
	we simply choose to use a hash table to store the configurations. 
This is done using Python's \texttt{OrderedDict} implementation. 
However, because we enforce local symmetries ($\hat{N}$ and $\hat{S}_z$),
	we use a nested hash table. 
By specifying a Fock space configuration over $N$ clusters as an immutable tuple-of-tuples 
\texttt{((N$_\uparrow^1$,N$_\downarrow^1$), (N$_\uparrow^2$,N$_\downarrow^2$), $\dots$,(N$_\uparrow^N$,N$_\downarrow^N$) )}
all TPS states with the same distribution of particle numbers can be described with a tuple of state indices:
\texttt{$(\alpha, \beta, \dots, \gamma)$}.
Consequently, an arbitrary TPS expansion coefficient can be accessed by two sequential hash table lookups:
\begin{align}\tt
	&\tt TBL2 = TBL1[((N_\uparrow^1,N_\downarrow^1), (N_\uparrow^2,N_\downarrow^2), \dots, (N_\uparrow^n,N_\downarrow^n))] \\
	&\tt c(\alpha^{N_\uparrow^1,N_\downarrow^1}, \beta^{N_\uparrow^2,N_\downarrow^2}, \dots, \gamma^{N_\uparrow^n,N_\downarrow^n}) =  
	 TBL2[\alpha,\beta,\dots,\gamma] 
\end{align}
This approach seems quite appropriate for dealing with variational degrees of freedom, of which there are typically only a modest number (i.e., fewer than around 100k).
However, when computing the first-order perturbative correction to the wavefunction, this number of configurations can create hash tables too large to store in memory. 
We have addressed this problem in the short term, by simply (but at significant CPU work) pruning small values before adding them to our hash table,
	using $\epsilon_s$ from Eq. \ref{eq:epsilon_s}.
However, a better solution would be to devise a low-memory PT2 correction for a TPS basis analogous to either the deterministic approach 
in Ref. \onlinecite{tubmanEfficientDeterministicPerturbation2018} or the semistochastic approach in Ref. \onlinecite{sharmaSemistochasticHeatBathConfiguration2017}. 

In addition to the storage and manipulation of the state vector information, 
storing the operator tensors from Eq. \ref{eq:mel1} also requires some care. 
Our code is organized in a class-based structure, such that each cluster is an instance of a \texttt{Cluster} class, 
which owns all local operator tensor data, stored as dense arrays.
However, because we have preserved local symmetries, we can reduce the storage by keeping only the operator 
transitions which are not symmetry forbidden. 
As such, in order to access an operator tensor, we again use a hash table to map specific Fock space transitions to operator tensors. 
For example, consider two states, \texttt{$\ket{\gamma}_I$} and \texttt{$\ket{\delta}_I$}, 
	which live in Fock spaces $\tt{(N_\uparrow^I,N_\downarrow^I)}$ 
	and $\tt{({N'}_\uparrow^I,{N'}_\downarrow^I)}$, respectively.
The operator tensors associated with the $\hat{p}^\dagger\hat{q}^\dagger\hat{r}$
operators only have data available if symmetry allowed, i.e., if $\tt N_\uparrow^I == {N'}_\uparrow^I+1$ and $\tt N_\downarrow^I == {N'}_\downarrow^I$.
If that's satisfied, then the dense tensor can be retrieved by a hash table lookup taking in the local Fock space transition:
\begin{align}
	\Gamma_{pqr}^{\gamma\delta} = \tt DATA[string(AAa)][(N_\uparrow^I,N_\downarrow^I,{N'}_\uparrow^I,{N'}_\downarrow^I)]
\end{align}
where \texttt{string(AAa)} indicates a request for an operator string with three $\alpha$ operators (\texttt{A}=$\alpha$ spin,  \texttt{B}=$\beta$ spin), 
and the first two are creation (upper case) and the last is annihilation (lower case).
This allows us to have a fine grained control over the tensor contractions used to form matrix elements, 
	preventing the computation of any hard zeros. 
Because the Hamiltonian contains up to four-index quantities, a naive strategy would also store the full set of two-particle transition densities, 
$\Gamma_{pqrs}^{\alpha,\alpha'}$.
However, because these terms can only contribute to local Hamiltonian terms, we can precontract these terms into a local Hamiltonian matrix, avoiding the need to store the 
6 index quantities, leaving the memory bottleneck to be the 5 index terms: $\Gamma_{pqr}^{\alpha\alpha'}$.
This memory bottleneck prevents us from considering exact clusters larger than 6 orbitals. 
The size of this tensor, $\Gamma_{pqr}^{\alpha\alpha'}$, is  
	$\mathcal{O}(N^3 M M')$ where $M$ is the number of states in the largest Fock space, $M'$ is the number of states in the next largest Fock space with one electron different,
	and $N$ is the number of orbitals in the cluster.
Because the number of states,  $M$,  increases factorially, 
	it is difficult to store data for more than a few clusters having 6 orbitals. 
One way to reduce the memory demands is to truncate the number of cluster states. 
This can be done either by energy or by entanglement measures, as is outlined in the Appendix.
Additional improvements can be made by manually handling the various tensor contractions.
Currently these  are handled in a rather abstract way  which  prevents much customization.

By the nature of the algorithm each non-trivial step is relatively easily paralellized. 
We have implemented the most expensive steps using shared memory parallelization, and have seen good scaling on the machines we've tested this on,
systems with 24 or 32 cores. However, it would be relatively straightforward to parallelize over many nodes, and we plan on doing this in the near future.

\subsection{Related works}
There are several  approaches in the literature which share the orbital clustering feature and  TPS representation used in TPSCI. 
Perhaps the work most closely related to TPSCI is the Block Correlated Coupled Cluster (BCCC) approach of Li and coworkers.\cite{fangBlockCorrelatedCoupled2007,fangBlockCorrelatedCoupled2008,fangBlockCorrelatedCoupled2008a,liBlockcorrelatedCoupledCluster2004,shenBlockCorrelatedCoupled2009,xuBlockCorrelatedSecond2013}
In BCCC, the orbitals are grouped into clusters and the wavefunction is represented in a TPS basis. 
Then, inter-cluster correlations are treated with an exponential parameterization, 
whose amplitudes are solved for non-linearly.
The excitonically renormalized CC (XR-CC) also works in a TPS basis where the state-to-state interaction is solved for in a CC fashion.\cite{Liu2019}
Our method shares the TPS basis, but differs in both the treatment of intercluster correlation, and  in the definition of the block states. 

Another related cluster-based approach is the Renormalized Exciton Method (REM)\footnote{One should be careful to not confuse this approach with the musical group from Athens, Georgia.}
of Malrieu and coworkers.\cite{AlHajj2005a,Zhang2012a,Ma2012,Ma2013}
In contrast to both TPSCI and BCCC, REM includes the intercluster interactions via a Bloch effective Hamiltonian.
In terms of the implementation, our approach is most closely related to the Active Space Decomposition 
(ASD) of Shiozaki and coworkers.\cite{Parker2013a}
The ASD method was extended to more than two clusters using a DMRG type wavefunction.\cite{Parker2014,Soichiro2019}
One can view TPSCI as a generalization of ASD to arbitrary numbers of clusters, 
with the global state optimization being approximated with CIPSI rather than the exact subspace diagonalization used in their work. 
Again, the Tucker  decomposition basis is another distinguishing  aspect of our current work.
Another approach which helped inspire our current work is the Cluster Mean Field (cMF) method of Scuseria and coworkers.\cite{Jimenez-Hoyos2015}
In fact, in most of the numerical calculations below, we use the fully optimized (both orbital and cluster state rotations) cMF as the reference TPS for TPSCI.
The cMF method has also been extended to ab initio systems with the name variational localized active space self consistent field (vLASSCF).\cite{hermes2020variational}
cMF and TPSCI also  share the  orbital clustering, but TPSCI goes beyond the variational description of a single TPS and also defines the Cluster states through a Tucker decomposition.
Other TPS methods include the rank-one basis proposed for 
	molecular aggregates,\cite{Soichiro2019}
	and the ab initio Frenkel Davydov Exciton Model (AIFDEM) of Herbert and co-workers for modeling 
	the low-lying singly excited states of aggregates using monomer direct product basis.\cite{Morrison2014}

The TPSCI method  also shares some features with other approaches which do not necessarily work with a TPS basis, 
but still involves some degree of  orbital clustering. 
The ORMAS (occupation restricted multiple active space) method restricts the  number of electrons
	in different orbital blocks and truncates the configuration expansion while still in the determinant basis.\cite{Ivanic2003}
The Restricted Active-Space (RAS) method \cite{Olsen1988} and Generalized Active-Space method
are also approaches which involve orbital clustering, but the similarities essentially end there.\cite{Ma2011}

\section{Results and Discussion}
The main goal of TPSCI is to make the calculation of large molecules possible when the number of determinants get intractable for determinant based selected CI.
Hence one of the main focuses in this paper will be a comparison between the TPSCI method and Slater determinant based SCI
	to understand how clustering impacts the compactness of the representation.  
We compare mainly two aspects, the accuracy vs.\ final dimension of the variational space.

\paragraph{How to compare compactness of different methods?}
There is indeed some ambiguity in deciding how to compare TPSCI with determinant based SCI methods. 
Ultimately, it's not immediately clear what should be treated as a variable when counting degrees of freedom. 
On the one hand, since TPSCI involves diagonalizing local Hamiltonians to define the cluster basis, one might consider these local wavefunction coefficients as variables. 
Thus adding to the parameter count for TPSCI. 
Similarly, we could add orbital coefficients into the list of variables for both TPSCI and SCI. 
On the other hand, one might prefer to define variables to be parameters optimized by a global objective function (full system energy).
In this paper we have chosen the latter definition, as this seems to be more consistent with the literature (e.g., basis set contraction coefficients aren't usually considered degrees of freedom in post-SCF calculations),
and because it is more closely related to the computational cost (the initial cMF calculation is much faster than the resulting TPSCI).
As such, throughout the results section, we will make comparisons between different methods based on the number of degrees of freedom which are optimized to minimize the full molecule's energy.
Thus the term ``Dimension'' will refer to the number of Slater determinants or TPS's.

In many of the results below, we use a simple convergence technique we refer to as ``bootstrapping'' which avoids going through larger dimensioned 
	intermediate steps during the Tucker optimization. 
This approach is explained in the Supplementary Information. 


In section \ref{sec:hubbard} we study a simple modified Hubbard model which allows us to manually tune the impact of ``clusterability'' on the performance of TPSCI.  
Because SCI methods were unable to accurately model the Hubbard model,
we use DMRG as a benchmark to compare the TPSCI results.
The DMRG results were obtained using the \texttt{ITensor} library.\cite{ITensor}
In section \ref{sec:diatomics} we study smaller ab initio systems.
We present data for N$_2$ bond dissociation curve with 6-31G basis, active space = (10e, 16o).
We also present data for cc-pVDZ basis set results for N$_2$ and F$_2$ molecules at their equilibrium and stretched bond lengths,
	having active spaces of (10e, 26o) and (14e, 26o), respectively.

Finally in section \ref{sec:pah} we study the ground state energies for a few $\pi$ conjugated systems.
The largest molecule in our test set is hexabenzocoronene, which has an active space of (42e, 42o).
The geometries were optimized using B3LYP/cc-pVDZ level of theory, 
	 and the {\tt xyz} files can be found in the Supplementary Information.
The HCI data quoted for the Hubbard model, N$_2$ molecule and the $\pi$ conjugated systems were obtained using the Arrow package.\cite{Holmes2016,Li2018,sharmaSemistochasticHeatBathConfiguration2017}
The PT correction used for HCI is computed semistochastically (SHCI).\cite{Li2018}
The ASCI data for the Hubbard model were generated using the \texttt{Q-Chem} package.\cite{Shao2014} 
The integrals for all the molecular examples were computed using \texttt{PySCF}.\cite{Sun2018}

\subsection{Hubbard Model}\label{sec:hubbard}

Model Hamiltonians provide a useful tool for exploring the behavior of different approximate simulation techniques. 
In this section, we use the Hubbard model to explore how the inherent ``clusterability'' of
	a system of fermions affects the performance of TPSCI. 

\begin{figure*}
	\includegraphics[width=\linewidth]{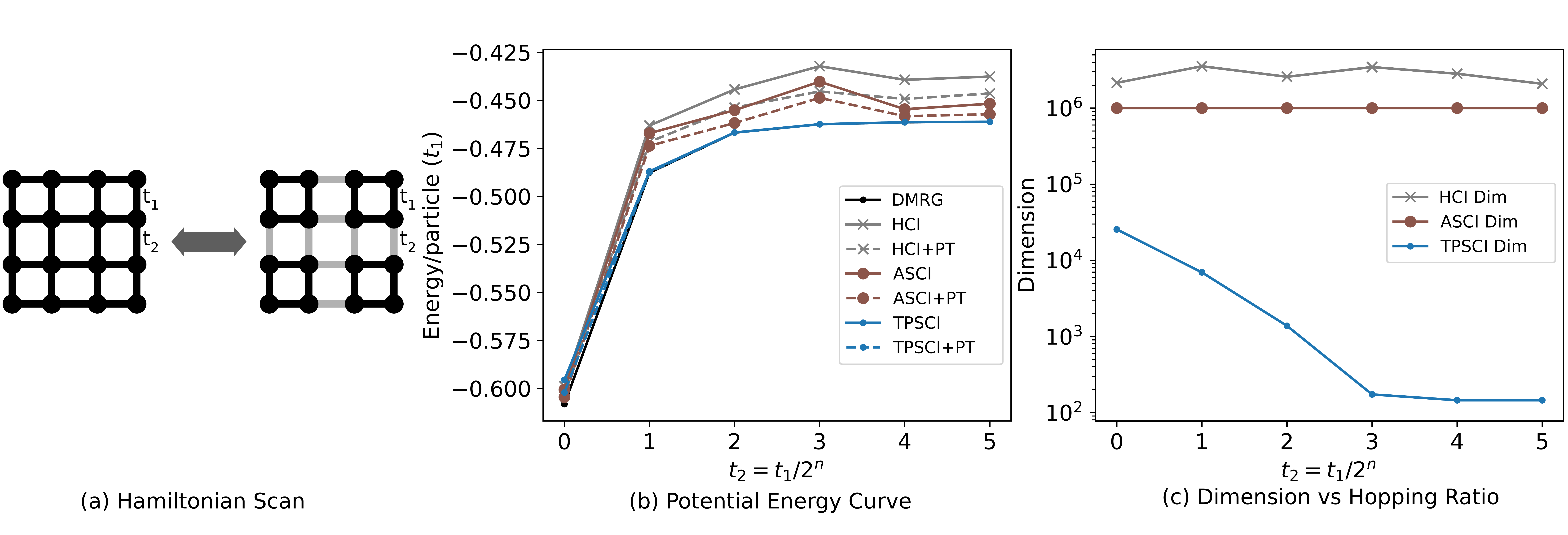}
	\caption{Clusterability of the Hubbard model. (a) Schematic representation of the Hubbard model used for the data, 
		dark lines correspond to $t_1$ and lighter lines correspond to $t_2$.
		(b) Energy/site of the system as the $t_1:t_2$ ratio is changed.
		TPSCI lines are nearly indistinguishable from the DMRG results.
		(c) Comparison of the dimension of the variational space as the $t_1:t_2$ ratio is changed.
		DMRG result uses M=1600.
		The TPSCI calculations reported use {\tt ($\epsilon$=5e-8 $\epsilon_c$=1e-2 $\epsilon_s$=1e-7)} with MP PT correction. 
		}
	\label{fig:data-hubbard-2d}
\end{figure*}
\subsubsection{Hubbard: Effect of Clusterability}
Being motivated by the assumption that one can find some local structure in the Hamiltonian, 
	the uniform Hubbard model is taken as our worst case scenario, and we expect our method to be inefficient in this domain.
Since we are working in a TPS basis, the best case scenario would be a Hamiltonian with no interactions between clusters. 
In such a case, the exact ground state becomes a single TPS. 
We expect a wide variety of physical systems to occur between these artificial limits. 
As such, we start in Fig. \ref{fig:data-hubbard-2d}(a) by exploring the transition between uniform lattice to a highly clusterable lattice,
by scanning the relative strength of the Hamiltonian coupling between clusters. 
The Hamiltonian used has two distinct hopping terms, and one electron-electron repulsion term:
\begin{align}
	\hat{H}=& \sum_I\sum_{\langle i,j \in I \rangle \sigma}-t_{1} c_{i\sigma}^{\dagger}
	c_{j\sigma}+\sum_{IJ}\sum_{\langle i \in I,j \in J \rangle \sigma}-t_{2}
c_{i\sigma}^{\dagger} c_{j\sigma} \nonumber \\
	&+U \sum_{j} n_{i \uparrow} n_{j \downarrow} 
\end{align}
where $t_1$ ($t_2$) denote hopping within (between) clusters, and $U$ is the Coulomb repulsion. 
In order to make this a strongly correlated system, we set $U=5t_1$ and start with a uniform lattice, $t_1=t_2=1$.
We then change the magnitude of of the inter-cluster hopping, $t_2$, scaled as $\frac{t_1}{2^{n}}$ where $n$ varies from 1 to 5.

In Fig. \ref{fig:data-hubbard-2d}, we observe confirmation that the accuracy and compactness of TPSCI should increase
	with increasing clusterabilty.
We compare TPSCI with two different determinant based SCI methods, the SHCI and ASCI methods. 
For this strongly correlated system, Slater determinant-based SCI methods were not able to find accurate results for any point on this scan,
	with reasonable numbers of variational parameters.
For the ASCI method, 1 million determinants were included in the variational space 
	while the SHCI results were computed using $\epsilon_1$ = {\tt 5e-4} and $\epsilon_2$ = {\tt 1e-9}.
For these systems, we present data with the HF basis since it gave better results compared to the local basis.

At higher ratios of $t_1:t_2$, the TPSCI results are almost exact with a variational space of less than a few thousand configurations.
This is a result of the fact that the exact ground state is moving increasingly close to a single TPS.
Hence in the single particle basis, the representation is not really sparse and
	far more determinants than computationally feasible might need to be included for such an example.
For large U/t ratio, the single particle basis would be even worse.
TPSCI on the other hand does not depend on this and hence can be used as a good alternative. 

\begin{figure}
	\includegraphics[width=\linewidth]{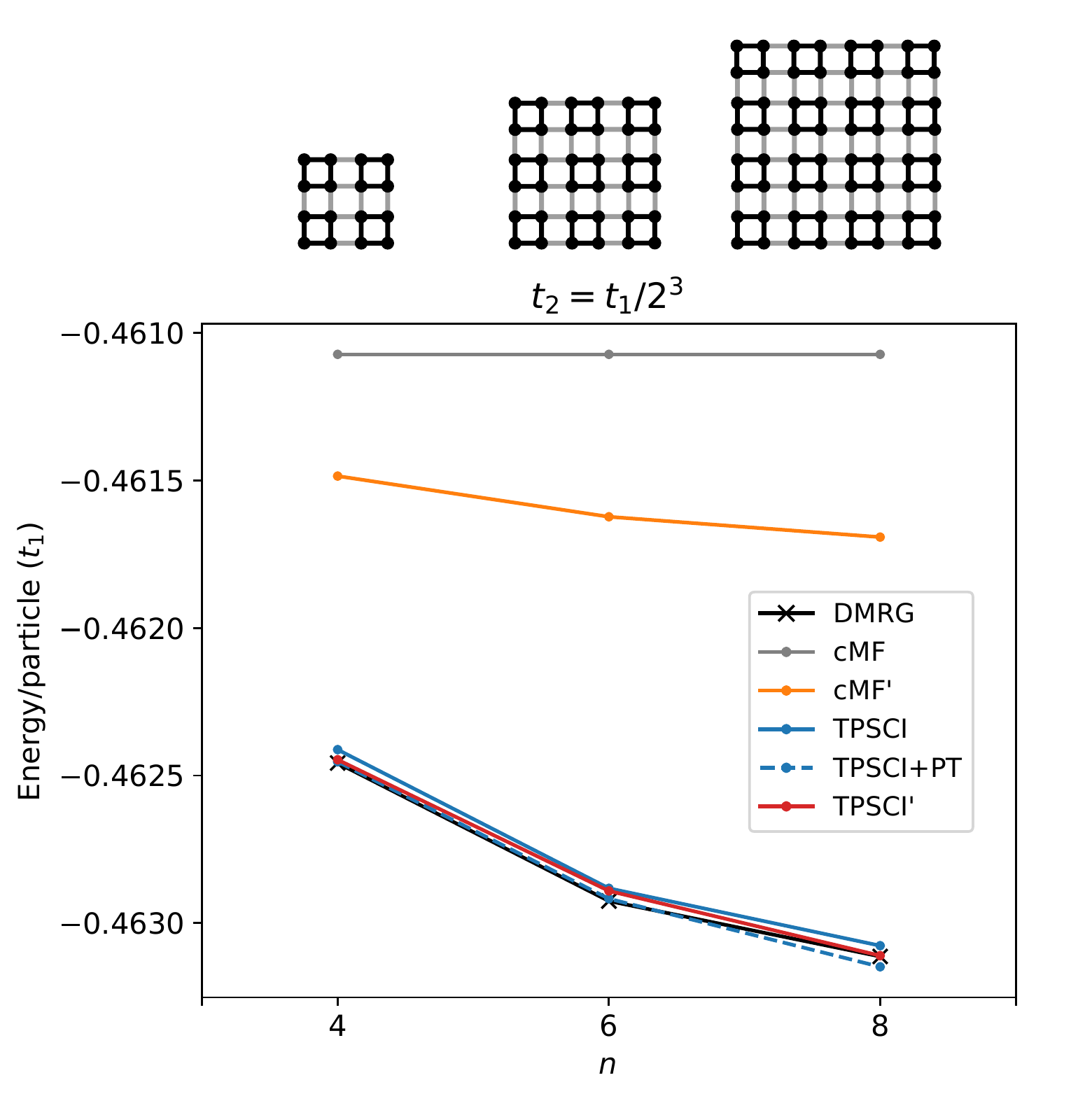}
	\caption{Size dependence of the Hubbard model is studied
	where we plot Energy/particle vs.\ lattice size.
	The intra-cluster:inter-cluster hopping ratio is fixed at  1/$2^3$.
	For CMF (CMF') orbitals are frozen (optimized).
	The TPSCI (TPSCI') dimensions for each data points are: 16-site: 173 (429), 36-site:  1073 (1205) , 64-site: 1978 (2735)
	The DMRG calculations were carried out with a 'snake-like' path to keep stronger interactions more local.\cite{Stoudenmire2012}
	}
	\label{fig:data-hubbard-2d-series}
\end{figure}

\subsubsection{Hubbard: Effect of Lattice Size}
In this second example, we explore how accuracy and dimension of the variational space changes when increasing the size of the system.
To do this, we fix the $t_1/t_2$ ratio to $\tfrac{1}{8}$.  
Consistent with the previous section, we set $U=5t_1$.
We start with the 16-site problem from above, but now increase the system size from 16, to 36, to 64 site lattices, all at half-filling and antiferromagnetic. 
Based on the performance of SCI on the 16 site problem, we did not attempt to compute the SCI energies for these larger lattices. 
Also consistent with the above section, the TPSCI calculations uses a clustering in which all $t_1$-coupled sites form a cluster. 
As such, the three different systems have 4, 9, and 16 clusters, respectively.
This is shown in Figure \ref{fig:data-hubbard-2d-series}.

We use both orbital optimized and frozen orbital cMF references for the TPSCI calculation. 
We denote the frozen orbital (orbital optimized) version as TPSCI (TPSCI') and the corresponding reference as cMF (cMF').
We compare both TPSCI and TPSCI' results with DMRG values with a fixed M value of 1600.
We also plot the reference TPS (cMF and cMF') energy for these systems for comparison. 
Despite these being 2D systems, DMRG works quite well, although for the larger lattices (especially the 64 site lattice), 
	the accuracy of TPSCI approaches that of DMRG.
We note though, that it's difficult to achieve a fair comparison of TPSCI and DMRG, 
	as both can, in principle, be systematically improved to get arbitrarily accurate results.
Nonetheless, for the 64-site example, the variational energy for TPSCI' (using 2735 variables) is comparable to DMRG with M=1600.
The Hubbard model in the frozen basis is extremely sparse and PT correction for this 64-site could be easily computed, since the repulsion is diagonal.
For the orbital optimized version this is not the case and hence we have only included the variational energy correction.


One challenge arises when studying the Hubbard model with different Hamiltonian parameters.
Because the Hamiltonian enters into the selection criterion for the TPSCI method (via the first order amplitudes),
	we find that the TPSCI threshold value $\epsilon$ does not yield consistent convergence behavior.
This means that the accuracy of the method cannot be directly linked to the selection criteria
	when modeling different Hamiltonians. 
While we only notice this problem with the Hubbard Hamiltonian, it is something we plan to investigate more in the future. 
One strategy would be to develop a TPSCI version of the $\Lambda$-CI method of Evangelista
for growing the $\mathcal{P}$ space,\cite{Evangelista2014}
which is designed to have better accuracy guaranteed.

\subsection{Molecular diatomics}\label{sec:diatomics}
\subsubsection{Nitrogen dissociation with 6-31G}
While model Hamiltonians are useful for artificially exploring the behavior of an approximation, 
the ultimate goal of our work is to produce an efficient method for ab initio molecular modeling. 
To understand the convergence behavior for molecular electronic structure, 
we start with a canonical example of a small strongly correlated system: N$_2$ dissociation. 

For the Hubbard model it was straightforward to form the clusters based on sparsity of the hopping term.
In contrast, for small molecular systems the clustering is less straightforward.
For a diatomic system like N$_2$ molecule, traditional single reference methods like CCSD provide good results at 
	shorter bond lengths, but fail when the bond is stretched. 
Six orbitals in the N$_2$ molecule become degenerate when bond is stretched and therefore RHF reference is not good enough to represent the molecule. 

In this section, we study the N$_2$ example with frozen 1s orbitals.
Even though N$_2$ is a triple bonded system, the interaction between these bonds are not as strong as the bond itself.
Hence, we can partition the orbitals in the triple bond (6e, 6o) into  three (2e, 2o) clusters.
This would mean putting the $\sigma$ and $\sigma^{*}$ bonds in a cluster, and the two $\pi$ and $\pi^{*}$ bonds in separate clusters. 
We are left with the lone pairs and we leave them in a separate cluster.
This is similar to a perfect pairing type cluster since each cluster is a pair of bonding/antibonding orbitals.\cite{Hunt1972,Beran2005}  
Hence the clustering would look like this in a minimal basis: ($\sigma_s$, $\sigma^*_s$),($\sigma_{pz}$, $\sigma^*_{pz}$), ($\pi_{px}$, $\pi^*_{px}$), ($\pi_{py}$, $\pi^*_{py}$) 

In the 6-31G basis, there are extra 3s and 3p basis functions on top of the minimal basis.
Hence we have a total of 16 orbitals.
We can put the extra orbitals in bonding/antibonding pair clusters, similar to a perfect pairing type clustering.
This makes a total of 8 clusters. 
We can also combine the orbitals of same angular momentum but different principle quantum numbers as one cluster.
This would lead to a clustering with 4 orbitals per cluster. 
We refer to these two types of clusterings as 8c and 4c:
\begin{itemize}
	\small
	\item 8c:(2s), (3s), (2p$_z$), (3p$_z$), (2p$_x$), (3p$_x$), (2p$_y$), (3p$_y$) 
	\item 4c:(2s, 3s), (2p$_z$, 3p$_z$), (2p$_x$, 3p$_x$), (2p$_y$, 3p$_y$) 
\end{itemize}
By defining the clusters in this way in 4c, dynamic correlation for each bond is included within a cluster.

We compare these two clustering choices against HCI with $\epsilon_1$={\tt 2e-4} (variational part) and $\epsilon_2$={\tt 1e-9}.
For TPSCI, we use cMF with frozen orbitals as the reference, 
	with a selection threshold of $\epsilon$={\tt 5e-8},\footnote{since TPSCI selects on the probability and HCI selects on the magnitude, these thresholds are related by a square root}
	and with EN perturbative correction.
The search space for each iteration was defined with $\epsilon_c$={\tt 1e-3} and a screening of the $\mathcal{Q}$ space couplings was set to $\epsilon_s$={\tt 1e-7}.
We have found that setting the search threshold $\epsilon_s\leq ${\tt 1e-7} consistently provides sub-mH accuracy.

\begin{figure*}
	\begin{center}
	\includegraphics[width=\linewidth]{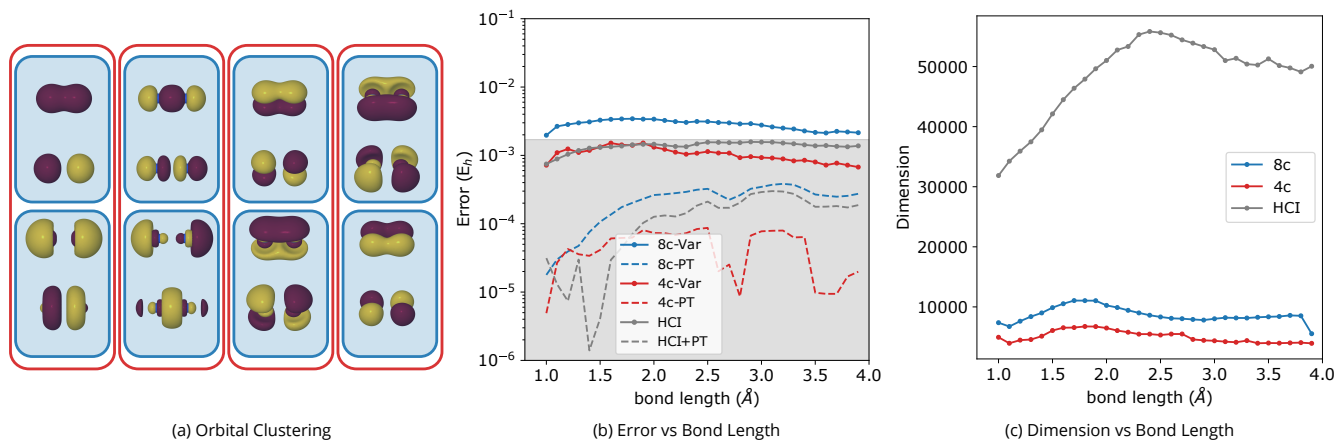}
	\caption{Nitrogen molecule with clustering based on bonding patterns. 
		(a) The molecular orbitals for N$_2$  and the clustering choices (4c) four clusters and (8c) eight clusters
		(b) Error with CAS-CI results for TPSCI method with the two different clustering options and HCI method.
		Grey area denotes regions with chemical accuracy, 1 kcal/mol.
		(c) dimension of the variational space along the PES scan
		The TPSCI calculations reported use {\tt ($\epsilon$=5e-8 $\epsilon_c$=1e-3 $\epsilon_s$=1e-7)} with EN perturbative correction. 
		The HCI calculations reported use {\tt ($\epsilon_1$=2e-4 $\epsilon_2$=1e-9}) with a semistochastic EN perturbative correction.
		}
	\label{fig:n2_pes}
	\end{center}
\end{figure*}

The molecular orbital clusterings, 4c and 8c, are pictorially depicted in Figure \ref{fig:n2_pes}(a).
In Figure \ref{fig:n2_pes}(b), we show the error with respect to FCI results.
The region of chemical accuracy is marked by 1 kcal/mol.
As seen from the figure \ref{fig:n2_pes}(b), the 8c results are not as accurate as the HCI or 4c results. 
These calculations can be made more accurate by using a lower threshold,
	but for clarity we chose to show data using same threshold for both 4c and 8c data.
The PT correction of the determinant based HCI is better at lower energies,
	and gets worse for the stretched geometries where the system is more strongly correlated.
The variational dimension for the HCI method increases as the bond is dissociated (more strong correlation).
This is in contrast to relatively constant dimension for TPSCI along the PES.

\begin{table*}[]
	\begin{center}
		\caption{Diatomic systems: N$_2$ and F$_2$ with cc-pVDZ basis. We compare both HCI and ASCI results with TPSCI. We also provide results with natural orbitals for ASCI calculations.
		The TPSCI calculations reported use {\tt ($\epsilon$=5e-8 $\epsilon_c$=5e-3 $\epsilon_s$=1e-7)}.
		The SHCI calculation use {\tt ($\epsilon_1$=5e-4 $\epsilon_2$=1e-9)} and all the ASCI calculations used a variational space of 50,000 determinants.
		*ASCI calculation landed on an excited state with natural orbitals.
		}
		\label{tab:dz_data}
			\begin{tabularx}{1.0\textwidth} {
                                                          >{\raggedright\arraybackslash}X
                                                         | >{\raggedright\arraybackslash}X
                                                          >{\raggedright\arraybackslash}X
                                                          >{\raggedright\arraybackslash}X
                                                          >{\raggedright\arraybackslash}X
                                                          >{\raggedright\arraybackslash}X
                                                          >{\raggedright\arraybackslash}X
                                                          >{\raggedright\arraybackslash}X
                                                          >{\raggedright\arraybackslash}X
                                                          >{\raggedright\arraybackslash}X }
			\hline\hline
			\multicolumn{1}{l|}{N$_2$}       & \multicolumn{1}{l}{Variational} & \multicolumn{1}{l}{PT2} & \multicolumn{1}{l|}{Dim}    & \multicolumn{1}{l}{Variational} & \multicolumn{1}{l}{PT2} & \multicolumn{1}{l|}{Dim}    & \multicolumn{1}{l}{Variational} & \multicolumn{1}{l}{PT2} & \multicolumn{1}{l}{Dim} \\ \hline
			\multicolumn{1}{l|}{r=1.0977} & \multicolumn{3}{c|}{r}                                                           & \multicolumn{3}{c|}{2r}                                                          & \multicolumn{3}{c}{3r}                                                       \\ \hline
			\multicolumn{1}{l|}{SHCI}     & -109.2692                & -109.2769               & \multicolumn{1}{l|}{37,577} & -108.9571                & -108.9668               & \multicolumn{1}{l|}{53,028} & -108.9477                & -108.9564               & 42,782                  \\
			\multicolumn{1}{l|}{TPSCI}    & -109.2694                & -109.2769               & \multicolumn{1}{l|}{8,274}  & -108.9607                & -108.9674               & \multicolumn{1}{l|}{15,659} & -108.9522                & -108.9572               & 12,744                  \\
			\multicolumn{1}{l|}{ASCI}     & -109.2723                & -109.2770               & \multicolumn{1}{l|}{50,000} & -108.9603                & -108.9673               & \multicolumn{1}{l|}{50,000} & -108.9515                & -108.9570               & 50,000                  \\
			\multicolumn{1}{l|}{ASCI-no}  & -109.2738                & -109.2771               & \multicolumn{1}{l|}{50,000} & -108.9456*               & -108.9512               & \multicolumn{1}{l|}{50,000} & -108.9524*               & -108.9568               & 50,000                  \\ \hline\hline
			\multicolumn{1}{l|}{F$_2$}       & \multicolumn{1}{l}{Variational} & \multicolumn{1}{l}{PT2} & \multicolumn{1}{l|}{Dim}    & \multicolumn{1}{l}{Variational} & \multicolumn{1}{l}{PT2} & \multicolumn{1}{l|}{Dim}    & \multicolumn{1}{l}{Variational} & \multicolumn{1}{l}{PT2} & \multicolumn{1}{l}{Dim} \\ \hline
			\multicolumn{1}{l|}{r=1.4119} & \multicolumn{3}{c|}{r}                                                           & \multicolumn{3}{c|}{2r}                                                          & \multicolumn{3}{c}{3r}                                                       \\ \hline
			\multicolumn{1}{l|}{SHCI}     & -199.0913                & -199.0992               & \multicolumn{1}{l|}{68,994} & -199.0489                & -199.0554               & \multicolumn{1}{l|}{67,434} & -199.0489                & -199.0549               & 65,476                  \\
			\multicolumn{1}{l|}{TPSCI}    & -199.0911                & -199.0991               & \multicolumn{1}{l|}{6,225}  & -199.0501                & -199.0556               & \multicolumn{1}{l|}{3,694}  & -199.0498                & -199.0551               & 3,956                   \\
			\multicolumn{1}{l|}{ASCI}     & -199.0923                & -199.0993               & \multicolumn{1}{l|}{50,000} & -199.0498                & -199.0556               & \multicolumn{1}{l|}{50,000} & -199.0499                & -199.0550               & 50,000                  \\
			\multicolumn{1}{l|}{ASCI-no}  & -199.0937                & -199.0994               & \multicolumn{1}{l|}{50,000} & -199.0500                & -199.0556               & \multicolumn{1}{l|}{50,000} & -199.0496                & -199.0551               & 50,000                  \\ \hline\hline
		\end{tabularx}
	\end{center}
\end{table*}

\begin{figure}
	\begin{center}
	\includegraphics[width=\linewidth]{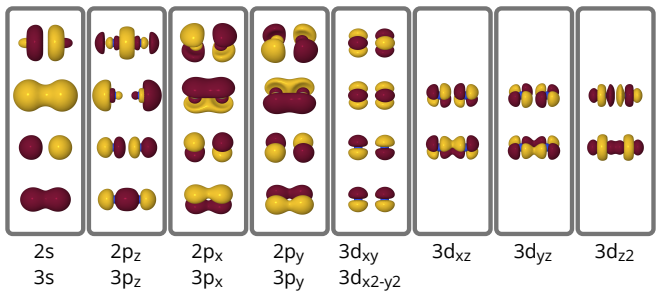}
		\caption{Choice of active-space orbital clustering for cc-pVDZ basis sets for N$_2$ and F$_2$}
	\label{fig:ccpvdz_orbs}
	\end{center}
\end{figure}

\subsubsection{Diatomics with cc-pVDZ basis}
In order to observe the impact of dynamical correlation, we have also performed TPSCI calculations for the larger basis-set, cc-pVDZ.
As in the previous subsection, we cluster with bonding/antibonding pairs. 
We consider three different bond lengths for both N$_2$ and F$_2$.

While we follow a similar clustering pattern as the 6-31G results above, 
in the cc-pVDZ basis, extra d-shell orbitals are present leading to more clusters.
To keep the size of the clusters small,
for the present paper, we partition the d-orbitals such that the atomic pairs coming from $d_{z^2}$, $d_{xz}$ and  $d_{xy}$ each form cluster. 
We leave the two $d_{xy}$ and $d_{x^2-y^2}$ as a 4-orbital cluster.
Taken together, this creates a total of 8 clusters for both N$_2$ and F$_2$.
By adding higher principle quantum number orbitals into the cluster, we are effectively allowing the clusters to become dynamically correlated. 
The clustering pattern is described below, keeping in mind that each orbital listed represents two orbitals, one from each atomic center.
\begin{itemize}
	\item (2s, 3s), (2p$_z$, 3p$_z$), (2p$_x$, 3p$_x$), (2p$_y$, 3p$_y$), (3d$_{xy}$, 3d$_{x^2-y^2}$), (3d$_{xz}$), (3d$_{yz}$), (3d$_{z^2}$)  
\end{itemize}
This clustering is shown for N$_2$ in Fig. \ref{fig:ccpvdz_orbs}
In Table \ref{tab:dz_data}, we present data for three bond lengths r, 2r, 3r where r=1.0977 (1.4119) for N$_2$ (F$_2$).
The thresholds for all the TPSCI calculations were {(\tt $\epsilon$=5e-8, $\epsilon_c$=5e-3, $\epsilon_s$=1e-7)},
	and the frozen cMF reference state was used. 
The HCI calculation were computed using $\epsilon_1$ = {\tt 5e-4}, 
	while the ASCI calculations were performed using 50K determinants. 
The most immediate conclusion, is that TPSCI converges to chemical accuracy with a smaller dimension than the determinant based methods. 
It has been previously shown that the use of natural orbitals can improve determinant based selected CI algorithms.\cite{Tubman2020,Holmes2017}
By using natural orbitals with ASCI, we observe similar behavior, although even with natural orbitals, the wavefunction in determinant based SCI is not as compact as the TPSCI wavefunction.
The use of natural orbitals seems to have the largest effect near equilibrium bond distances, with a smaller dependence on orbitals occuring at stretched distances.

\begin{figure*}
	\begin{center}
	\includegraphics[width=\linewidth]{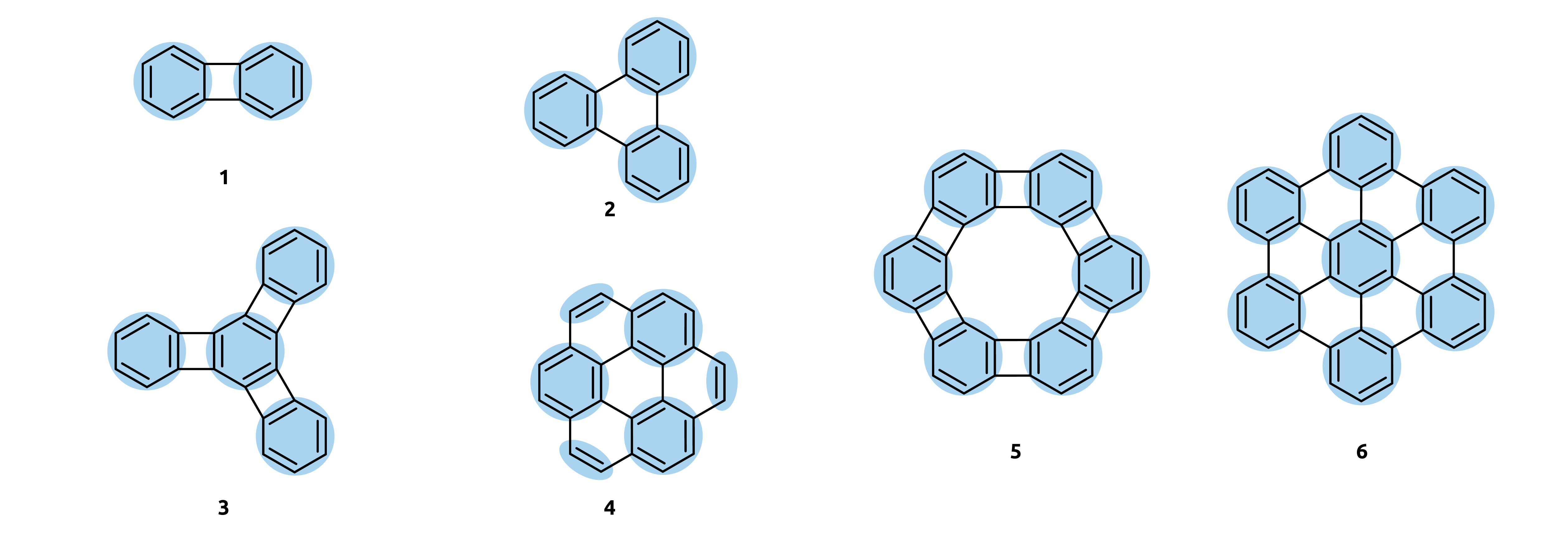}
		\caption{$\pi$-conjugated systems for TPSCI calculation with blue circles representing the clusters for the system.}
	\label{fig:pi-data}
	\end{center}
\end{figure*}

\begin{table*}[]
	\caption{Comparison between TPSCI and HCI for the $\pi$-conjugated systems used in the study. 
		The molecules are labelled according to Figure \ref{fig:pi-data}.
		The TPSCI calculations reported use {\tt ($\epsilon$=1e-7 $\epsilon_c$=1e-2 $\epsilon_s$=1e-6)}. 
		The HCI calculations use {\tt ($\epsilon_1$=1e-5  $\epsilon_2$=1e-9)} for molecules \textbf{1}-\textbf{4}.
		For \textbf{5} ({\tt $\epsilon_1$=3e-5 $\epsilon_2$=1e-9}) and for \textbf{6} ({\tt $\epsilon_1$=4e-5 $\epsilon_2$=1e-9}).
		}
	\label{tab:pi-data}
	\begin{tabularx}{1.0\textwidth} {
				  >{\raggedright\arraybackslash}X
				| >{\centering\arraybackslash}X
				  >{\centering\arraybackslash}X
				| >{\centering\arraybackslash}X
			 	  >{\centering\arraybackslash}X
				| >{\centering\arraybackslash}X
				  >{\centering\arraybackslash}X
				| >{\centering\arraybackslash}X
				  >{\centering\arraybackslash}X }
	\hline\hline
	molecule             & \multicolumn{2}{c|}{Variational} & \multicolumn{2}{c|}{PT2}  & \multicolumn{2}{c|}{Extrapolated}      &  \multicolumn{2}{c}{Dim} \\ 
	 	             & TPSCI      & SHCI       & TPSCI      & SHCI       & TPSCI 	             & SHCI 	    &  TPSCI   & SHCI       \\\hline 
	\textbf{1}           & -453.6310  & -453.6310  & -453.6310  & -453.6310  & -453.6310                 & -453.6310    &  201     & 174,757    \\
	\textbf{2}           & -680.5951  & -680.5906  & -680.5958  & -680.5944  & -680.5964                 & -680.5970    &  2,440   & 7,397,514  \\
	\textbf{3}           & -904.9121  & -904.8865  & -904.9136  & -904.9012  & -904.9142                 & -904.9144    &  3,885   & 21,179,338 \\
	\textbf{4}           & -905.2157  & -905.2100  & -905.2238  & -905.2212  & -905.2307                 & -905.2307    &  16,272  & 19,510,272 \\
	\textbf{5}           & -1353.8138 & -1353.6927 & -1353.8199 & -1353.7509 & -1353.8244                & -1353.8259   &  10,376  & 20,232,920 \\
	\textbf{6}           & -1582.4291 & -1582.2378 & -1582.4402 & -1582.3255 & -1582.4482                & -1582.4396   &  20,325  & 11,194,996 \\ \hline\hline
	\end{tabularx}
\end{table*}

\begin{figure*}
	\begin{center}
	\includegraphics[width=\linewidth]{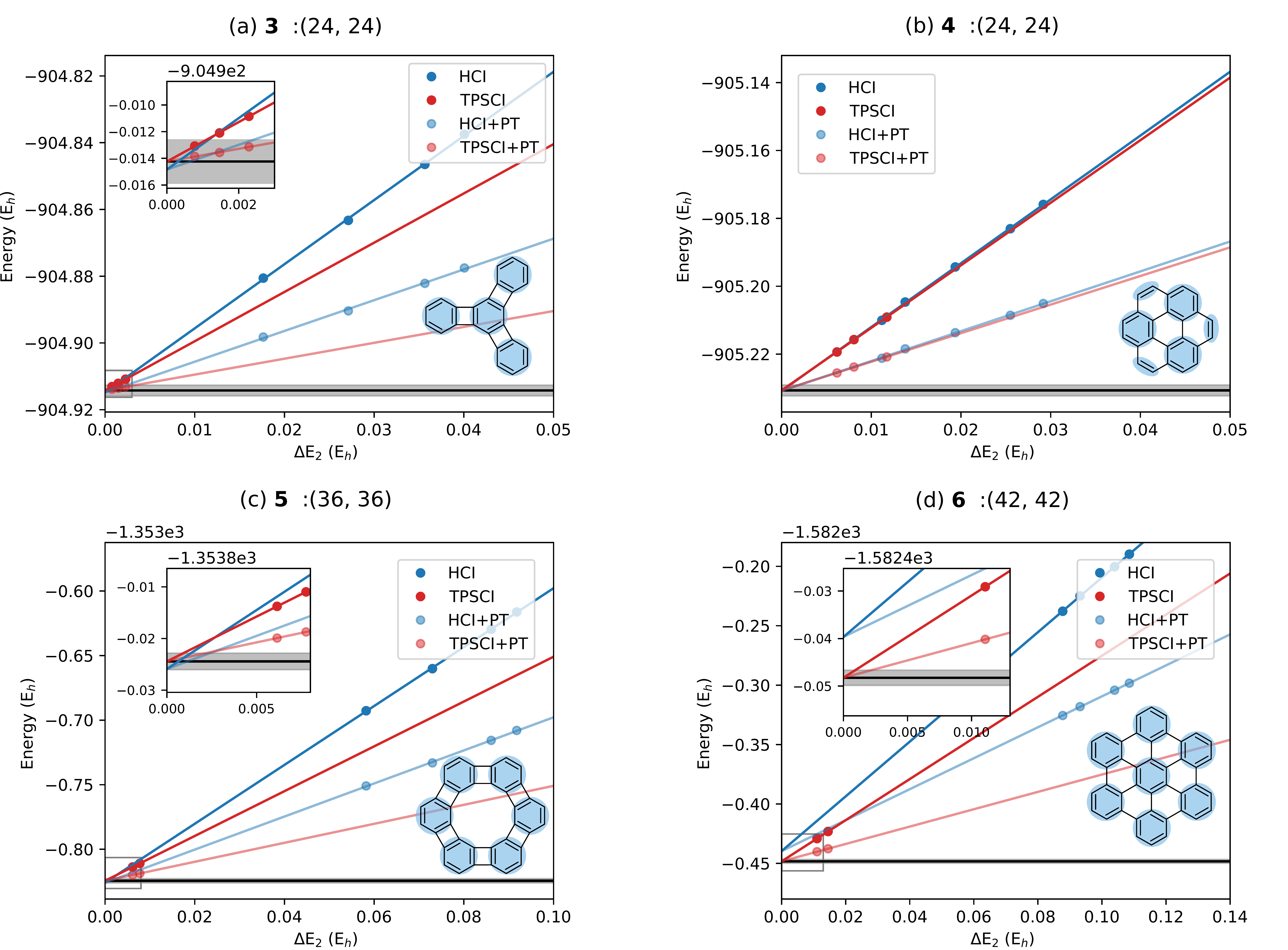}
		\caption{Extrapolation scheme for the four large molecules considered. The grey area corresponds to $\pm$ 1 kcal/mol about the extrapolated TPSCI energy (solid black line)}
	\label{fig:extrap24}
	\end{center}
\end{figure*}

\subsection{$\pi$-conjugated systems}\label{sec:pah}
While the small diatomic molecules of the previous section provide a rigorous test of the TPSCI method for systems which are well described by existing SCI methods like HCI and ASCI,
our goal in designing TPSCI is to model active spaces which are larger and more strongly correlated than what can be afforded with the traditional Slater determinant-based approaches. 
For this reason, we now turn our attention to systems which are expected to be good applications of TPSCI, 
	systems which are non-linear (so a MPS is non-ideal), strongly correlated, and somewhat clusterable. 
Taking poly-aromatic hydrocarbons (PAH's) as test cases, might perhaps be surprising because the characteristic delocalization of the $\pi$ system might seem to suggest 
	exactly the worst case scenario for observing ``clusterability''. 
However, the success of Clar's rule in relating the number of disjoint benzene units with stability,
	seems to suggest that a clustering pattern chosen from Clar's rule might be effective.
A suggestion that is consistent with recent results using fragment-based DFT methods.\cite{Noffke2020}


We have chosen a few example PAH's (shown in Fig. \ref{fig:pi-data}) ranging in size and clusterability, 
	with the clustering pattern consistent with Clar's rule shown in blue.
In addition to the well-known coronene-type bonding pattern ({\bf 2, 4, 6}), 
	we have also considered a few systems which contain rather strained 4-center ring bridging units ({\bf 1, 3, 5}). 
Though less stable, materials based on this bonding pattern have recently been synthesized using polymerization reactions of  1,3,5-trihydroxybenzene molecules.\cite{du_new_2017}
These material can have interesting applications because of its planar structure like graphene and porous nature.
For the sake of simplicity, we use the minimal STO-3G basis for these systems, 
	since we are only studying the $\pi$-conjugated electrons.\footnote{We have also used the cc-pvdz basis set with the same size active space and found similar qualitative results, 
		but with consistently smaller amount of active-space correlation.}
The active space for molecules \textbf{1}-\textbf{6} consist of 12, 18, 24, 24, 36, 42 orbitals, respectively. 

To initialize the clusters, we started with localized orbitals 
	and then performed a cMF calculation to obtain a cluster basis, and optimal orbitals. 
We found that the orbital optimization (only mixing orbitals within the active space and between clusters) 
	lowered the energy a substantial amount compared to the frozen cMF (-0.38 au in the case of {\bf 6}).
To minimize the memory requirements, we used the embedded Schmidt trunction (EST) approach described in the appendix, for each of these systems
discarding Schmidt vectors with singular values smaller than {\tt 1e-4}. 
This value was found to provide large memory (and cpu) savings, without significantly impacting the accuracy, 
with our tests indicating that the error was below 1mH. 
For each of these systems, we then performed three increasingly tight TPSCI calculations defined by the following settings:
\begin{enumerate}\tt 
	\item ($\epsilon$=1e-6, $\epsilon_c$=1e-2, $\epsilon_s$=1e-6)
	\item ($\epsilon$=1e-7, $\epsilon_c$=1e-2, $\epsilon_s$=1e-6)
	\item ($\epsilon$=1e-8, $\epsilon_c$=1e-2, $\epsilon_s$=1e-6)
\end{enumerate}
each time performing an MP perturbative correction setting {\tt ($\epsilon_c$=0, $\epsilon_s$=1e-7)}.
Only a single Tucker iteration was performed for the {\tt $\epsilon$=1e-6} calculation, and then that basis was used for the more accurate calculations.
For molecules \textbf{5} and \textbf{6}, we were unable to obtain the PT2 correction for the {\tt $\epsilon$=1e-8} calculations.

For HCI and ASCI, we used canonical HF orbitals, as we found that natural orbitals didn't have a significant affect for these systems. 
The HCI was computed for molecule \textbf{1}-\textbf{4} using {\tt ($\epsilon_1$=1e-5, $\epsilon_2$=1e-9)}.
For molecule \textbf{5} and \textbf{6}, the variational space became too large,
and thus the tightest data we could obtain for \textbf{5} and \textbf{6}  was {\tt $\epsilon_1$=3e-5}  and {\tt $\epsilon_1$=4e-5}, respectively.

In Table \ref{tab:pi-data}, we present the most accurate HCI data,
	alongside TPSCI data using the intermediate threshold level: {\tt ($\epsilon$=1e-7 $\epsilon_c$=1e-2 $\epsilon_s$=1e-6)}.
For biphenylene, \textbf{1}, we can see that both HCI and TPSCI give essentially the exact result for this small active space of (12e, 12o).
Nonetheless, even for this small system, the TPS basis is much more compact than a determinantal basis 201 vs. 174,757.
However, this shouldn't be too surprising, because \textbf{1} only has two clusters, meaning the ``Tucker'' decomposition performed at {\tt $\epsilon$=1e-6}
is actually an SVD, which is formally the most compact representation. 
For larger systems though, this is not the case, and the Tucker decomposition no longer presents a diagonal representation. 

Considering the larger molecules we continue to see a similar increase in compactness by around 3 orders of magnitude.
Molecules \textbf{2} and \textbf{3} collect only a few thousand configurations while still being more accurate than
	the HCI results with millions of determinants (Table \ref{tab:pi-data}).

\paragraph{Clusterability}Considering molecules \textbf{3} and \textbf{4} side-by-side, provides insight into the impact of clusterability. 
Both systems have the same active space size, (24e, 24o), but they differ in the connectivity of the clusters. 
Molecule \textbf{3} is able to be clustered into four 6-site clusters, a complete Clar's tiling. 
Molecule \textbf{4} on the other hand can only be grouped into three 6-site clusters, and three remaining 2-site clusters.
This has a significant impact on the compactness, with \textbf{4} requiring over four times the number of TPS's and being further from the extrapolated result.
Nonetheless, comparison with determinant based SCI is still impressive.
With approximately 16 thousand TPS configurations, the TPSCI energy for \textbf{4} is significantly lower than the determinant based SCI result with 19 million determinants.


\paragraph{Extrapolation}
Because of the slow convergence of correlation energy with variational dimension, 
it is often useful to use a few SCI calculations and extrapolate to the exact result.
We can also do this for TPSCI, provided each of the variational and PT2 energy pairs are computed using the same cluster basis.
In Fig. \ref{fig:pi-data}, we use the available energies to extrapolate to the exact result, 
giving us both an estimate of the FCI energy, and also an estimate of how accurate our TPSCI calculations are. 
TPSCI extrapolations are performed using all 3 accuracy levels, except for molecules \textbf{5} and \textbf{6}, for which only the first two calculations were used to extrapolate.
Comparing these extrapolations to the HCI extrapolations we observe that even though the TPSCI energies are much closer to converged, 
	the HCI extrapolations (with the exception of the largest system \textbf{6}) are very effective and provide comparable FCI estimates.
Overall, these systems indicate that the TPSCI method provides a unique representation able to more efficiently captures correlation in clusterable systems,
	extending the applicablity of SCI algorithms to larger systems.

\section{Conclusion and Future Work}
In this work, we have introduced a new selected CI method using tensor products of cluster states as the basis.
By folding the most important correlations into the basis vectors themselves, 
	much more compact wavefunctions can be obtained using the basic selected CI procedures, 
	a feature which can significantly improve the performance for strongly correlated systems. 
In choosing the nature of the cluster states, 
	we found that the Tucker decomposition provided a simple and efficient way to significantly improve the compactness of the 
final TPSCI wavefunction. 
Although our current code is far from optimized, we have demonstrated advantages over determinant based methods for large active spaces in PAH's. 
However, even if our implementation was sufficiently optimized, it's not obvious that TPSCI will provide a faster ``time to solution'' compared to 
	methods like HCI or ASCI for small systems in large basis sets due to the lack of clusterability. 
The TPS representation involves quite a bit of computational overhead, 
	which is only likely to pay off for spatially extended systems (like PAH's) which provide better opportunities for effective clustering. 

This initial paper presents the algorithm, an implementation, and proof of concept results.
However, much work remains to be done. 
A few of the ongoing and future works will involve:
\begin{enumerate}
	\item Develop efficient and low-memory PT2 correction algorithm, similar to that done in Ref. 
		\onlinecite{tubmanEfficientDeterministicPerturbation2018}.
	\item Investigating the behavior of different orbital clusterings. 
		Good performance of TPSCI requires the sensible partitioning of the orbitals into clusters. 
		In this work, we've largely done this by hand.
		However, this does not necessarily provide the best clustering.
		We plan on developing automated procedures for orbital clustering. 
		Preliminary results suggest that partitioning clusters based on the exchange matrix seems to work quite well.
		However, more rigorous approaches based on information theory\cite{Elvira2017,Luzanov2010,Stein2017} might provide improved results. 
	\item Exploring the performance of cluster basis truncation. 
		We have explored two different approaches (energy based, and entanglement based), 
		but it's not yet obvious what the best approach is, or how the technique depends on the chemical system. 
	\item Improved implementation. While our current code is efficient enough to obtain all the results in this manuscript, 
		including the (42e, 42o) results and the 64 site Hubbard lattice example, 
		many steps are far from optimal, with obvious ``hot spots'' occuring in pure Python functions. 
		Using C++ to reimplement these steps should provide significant improvements.
	\item Exploring how the different components of the various SCI methods such as HCI, ACI, ASCI, etc. behave in the TPS basis.
	\item Extending the method to study excited states. One possible route would be to use similar approach taken
		in the original CIPSI work. 
		Another interesting direction is to form a reference state using single excitations in clusters as done in AIFDEM\cite{Morrison2017}   
		and include important configurations avoiding collapse to the ground state.
		TPSCI would be especially well suited for modeling excitations in molecular aggregates,
		due to the lack of covalent bonds between systems making the TPS representation converge extremely quickly. 
\end{enumerate}

\section{Acknowledgements}
The authors would like to thank Garnet Chan for helpful suggestions regarding DMRG calculations.
This research was supported by the  National Science Foundation (Award No. 1752612).

\appendix
\section{Truncation of cluster states using approximate Schmidt vectors}
The TPSCI method described in the main text presents an algorithm for systematically approximating the exact solution within a basis of tensor products of many-body cluster states. 
This algorithm is quite general and can be used with cluster states obtained in a number of ways. 
The most straightforward approach would be to simply diagonalize the Hamiltonian operator which only acts locally on the cluster (i.e., all indices correspond to orbitals within the cluster). 
This has the convenience that local operators become diagonal, but lacks all interactions with other clusters. 
A better approach, would be to use the cluster mean-field (cMF) method developed by Jim\'enez-Hoyos and Scuseria.\cite{Jimenez-Hoyos2015}
Ignoring the orbital optimization component of cMF first, cMF is the variational minimization of a single tensor product state wavefunction. 
This amounts to defining each cluster's ground state as the lowest energy eigenstate of an effective local Hamiltonian, which includes the mean-field interactions with all clusters.
Because the energy of a single TPS only depends on each cluster's ground state, the higher energy cluster states aren't determined rigorously by a variational principle
as the cMF energy has a rotational invariance among the non-ground cluster state (this is analogous to problems in trying to tie physical meaning to virtual orbitals in Hartree-Fock).
However, the cMF local Hamiltonian is uniquely defined, and its associated eigenbasis (i.e., the ``canonical'' cluster state basis) provides an improved set of vectors for defining the cluster basis used in TPSCI. 

For small clusters (e.g., less than about 5 spatial orbitals) the full cluster state basis can be used without any complications. 
However, for larger clusters, the memory requirements needed to store the local operator tensors, $\Gamma_{p^\dagger qr}^{I_{\alpha},I_{\alpha'}}$, become significant. 
For example, the largest sector of Fock space for a six site cluster (3 $\alpha$ and 3 $\beta$ electrons) has a dimension of 400.
Storing the associated tensors requires about 2.5Gb per cluster: This doesn't include the numerous smaller particle number spaces. 
However, for a system that is clusterable one should be able to discard many of these states without significantly impacting the global ground state. 

Using the eigenvalue of the cMF Hamiltonian is one way to determine which cluster states to discard. 
However, because local energies aren't good predictors of entanglement, rather large numbers of states are needed to maintain accurate results for global quantities. 
To address this issue, we have developed a relatively simple approach for defining more compact cluster states, which we refer to as ``embedded Schmidt truncation'' (EST). 
For cluster $I$, the goal is to obtain a compact set of vectors which captures as much of its entanglement with the rest of the system as possible. 
The ideal set of vectors would then be obtained by diagonalizing the cluster $I$'s reduced density matrix obtained from the exact global system's ground state.
To develop a practical approximation to this, we instead decide to find the ground state of a smaller system comprised of cluster $I$ and a small number of bath orbitals which directly interact with $I$,
	using a mean-field description of the remaining system.
This approach is based on the ``Concentric Localization'' concept recently used in projection based embedding,\cite{claudinoSimpleEfficientTruncation2019a}
	and also density matrix embedding theory.\cite{kniziaDensityMatrixEmbedding2012}

In order to identify the bath orbitals for cluster $I$, we simply SVD the off-diagonal block of the exchange matrix between orbitals in cluster $I$ and orbitals in clusters $J\neq I$.
\begin{align}
	K^I_{pq} =& \sum_{LM}\sum_{r\in L}\sum_{s\in M}(pr|sq)D_{rs} &  \forall  p \in I,  q \notin I\\
		 =& \sum_s U^I_{p,s}\lambda_sV^I_{q,s} 
\end{align}
where $D_{rs}$ is the one-particle reduced density matrix resulting from the cMF calculation. 
The bath orbitals for cluster $I$ are defined to be those which directly interact with cluster $I$ via the exchange operator, 
and are trivially obtained by rotating the MO coefficients, $C_{\mu,p}$, by the right singular vectors:
\begin{align}
	C^{\text{bath},I}_{\mu,s} =& \sum_{J\neq I}\sum_{q\in J} C_{\mu,q}V^{I}_{q,s}
\end{align}
This creates a natural compression of exchange interactions, and the number of non-zero singular values, $\lambda_s$, is bounded by the number of orbitals in cluster $I$. 
The molecular orbitals are now organized into three subspaces, 
\begin{align}
	\mathbf{C}^I = \mathbf{C}^{\text{cluster,}I}|\mathbf{C}^{\text{bath,}I}|\mathbf{C}^{\text{env,}I}, 
\end{align}
 	where $C^{\text{env,}I}$ are the orbitals associated with the null space of the exchange coupling. 
This procedure can be continued recursively to define another bath, which completely captures the interaction between the previous bath and the environment, 
	creating a recursive approach to organizing all the orbitals by their ``nearness'' to cluster $I$.
In this work, we only consider the first layer bath.
The goal is now to obtain the exact solution to the combined cluster+bath system and then SVD the resulting ground state to define a basis of many body states to use for cluster $I$.
Because the bath orbitals are defined via the SVD, the number of bath orbitals will always be less than or equal to the number of orbitals in that cluster. 
For instance, if a cluster has 6 orbitals, then to obtain the basis, one would need to compute only the ground state for a 12 orbital problem. 
For larger clusters, this will quickly become expensive, but since only the ground state is needed, conventional selected CI algorithms like ASCI or SCI could be used alternatively for this step.

If the environment orbitals were unentangled with the cluster and bath orbitals, then the 1RDM in the environment space would be idempotent and we could simply perform a CASCI calculation to obtain this ground state. 
Because this is not the case generally, we simply purify the density in the environment space, and use this background density for the CASCI core.
Once the ground state of the embedded cluster$|$bath system is obtained, 
We organize the resulting CI vector into contributions to local particle number spaces and then perform an SVD. 
This provides us with a set of vectors with well defined particle numbers for cluster $I$, ordered according to their weight in the embedded ground state. 
This process is repeated separately for each cluster.

As described so far, this is simply a change of basis for the cluster states on $I$. 
No approximation has been made, and the final TPSCI wavefunction can still converge to the exact ground state. 
However, because our vectors are now weighted according to an approximate entanglement metric instead of energy, 
we can now perform a more aggressive truncation on the number of cluster states used to determine the dimension of the Hilbert space accessible to the TPSCI algorithm. 
In practice, this works extremely well when the system is well localized (e.g., polyaromatic hydrocarbons),
	and deleting cluster states with singular values smaller than .0001 seems to consistently have sub milliHartee impact on the results. 
Of course, for some systems this might not be ideal, and alternatives might need to be considered. 
This will be a focus of future work.


\providecommand{\latin}[1]{#1}
\makeatletter
\providecommand{\doi}
  {\begingroup\let\do\@makeother\dospecials
  \catcode`\{=1 \catcode`\}=2 \doi@aux}
\providecommand{\doi@aux}[1]{\endgroup\texttt{#1}}
\makeatother
\providecommand*\mcitethebibliography{\thebibliography}
\csname @ifundefined\endcsname{endmcitethebibliography}
  {\let\endmcitethebibliography\endthebibliography}{}

\end{document}